
\input harvmac

 1
 1
 1

\font\tenbf=cmbx10
\font\tenrm=cmr10

\font\eightrm=cmr8
\font\eightit=cmti8

\parindent=1.2pc
\magnification=\magstep1
\hsize=6.0truein
\vsize=8.6truein

\def\np#1#2#3{{\it Nucl. Phys.} {\bf B#1} (#2) #3}
\def\pl#1#2#3{{\it Phys. Lett.} {\bf #1B} (#2) #3}
\def\prl#1#2#3{{\it Phys. Rev. Lett.} {\bf #1} (#2) #3}
\def\physrev#1#2#3{{\it Phys. Rev.} {\bf D#1} (#2) #3}

\def\prep#1#2#3{{\it Phys. Rep.} {\bf #1} (#2) #3}

\def\cmp#1#2#3{{\it Comm. Math. Phys.} {\bf #1} (#2) #3}

\def\Tr{{\rm Tr}~}
\def\ev#1{\langle#1\rangle}
\font\litfont = cmr6

\def\half{{\litfont {1 \over 2}}}
\def\CM{{\cal M}}
\def\bZ{{\bf Z}}

\def\lL{{\cal L}}
\def\lfm#1{\medskip\noindent\item{#1}}
\overfullrule=0pt

\def\Im{{\rm Im ~}}
\def\CW{{\cal W}}
\def\half{{\litfont {1 \over 2}}}

\lref\sem{N. Seiberg, hep-th/9411149 , \np{435}{1995}{129}.}%
\lref\ils{K. Intriligator, R.G. Leigh and N. Seiberg, hep-th/9403198,
\physrev{50}{1994}{1092}.}%
\lref\intin{K. Intriligator, hep-th/9407106,
\pl{336}{1994}{409}.}%
\def\ilsii{\refs{\ils , \intin}}
\lref\isprev{K. Intriligator and N. Seiberg, hep-th/9506084, to appear
in the Proc. of Strings '95}
\lref\sv{M.A. Shifman and A. I. Vainshtein, \np{277}{1986}{456};
\np{359}{1991}{571}}
\lref\nonren{N. Seiberg, hep-ph/9309335, \pl{318}{1993}{469}}%
\lref\nati{N. Seiberg, hep-th/9402044, \physrev{49}{1994}{6857}}%
\lref\nonren{N. Seiberg, hep-ph/9309335, \pl{318}{1993}{469}.}
\lref\mo{C. Montonen and D. Olive, \pl {72}{1977}{117}; P. Goddard,
J. Nuyts and D. Olive, \np{125}{1977}{1}.}
\lref\dualnf{H. Osborn, \pl{83}{1979}{321}; A. Sen, hep-th/9402032,
\pl{329}{1994}{217}; C. Vafa and E. Witten, hep-th/9408074,
\np{432}{1994}{3}.}
\lref\swi{N. Seiberg and E. Witten, hep-th/9407087,
\np{426}{1994}{19}.}%
\lref\intse{K. Intriligator and N. Seiberg, hep-th/9408155,
\np{431}{1994}{551}.}%
\lref\swii{N. Seiberg and E. Witten, hep-th/9408099,
\np{431}{1994}{484}.}
\lref\ads{I. Affleck, M. Dine and N. Seiberg, \np{241}{1984}{493};
\np{256}{1985}{557}.}%
\lref\cern{D. Amati, K. Konishi, Y. Meurice, G.C. Rossi and G.
Veneziano, \prep{162}{1988}{169} and references therein.}%
\lref\kaplou{V. Kaplunovsky and J. Louis, \np{422}{1994}{57}.}%
\lref\finnpou{D. Finnell and P. Pouliot, RU-95-14, SLAC-PUB-95-6768,
hep-th/9503115.}%
\lref\vy{G. Veneziano and S. Yankielowicz \pl{113}{1982}{231};
T. Taylor, G. Veneziano and Yankielowicz, \np{218}{1983}{439}.}%
\lref\higgscon{L. Susskind, unpublished;
T. Banks, E. Rabinovici, \np{160}{1979}{349};
E. Fradkin and S. Shenker, \physrev{19}{1979}{3682}.}
\lref\rusano{V. Novikov, M. Shifman, A. Vainshtein and V.
Zakharov, \np{229}{1983}{381}}
\lref\isson{K. Intriligator and
N. Seiberg, hep-th/9503179,\np{444}{1995}{125}.}%
\rightline{hep-th/9509066, RU-95-48, IASSNS-HEP-95/70}
\vglue0.3cm
\centerline{\tenbf LECTURES ON  SUPERSYMMETRIC GAUGE THEORIES}
\baselineskip=18pt
\centerline{\tenbf AND ELECTRIC-MAGNETIC DUALITY\footnote{*}{\rm
To appear
in the Proc.\ of Trieste '95 spring school, TASI '95, Trieste
'95 summer school, and Cargese '95 summer school.}}
\baselineskip=18pt
\centerline{\eightrm K. INTRILIGATOR$^{1,2}$ and N. SEIBERG$^{1,2}$}
\baselineskip=18pt
\centerline{\eightit $^1$Department of Physics, Rutgers University}
\baselineskip=10pt
\centerline{\eightit Piscataway, NJ 08855-0849, USA}
\baselineskip=12pt
\centerline{\eightit $^2$Institute for Advanced Study}
\baselineskip=10pt
\centerline{\eightit Princeton, NJ 08540, USA}

\vglue0.4cm
\centerline{\eightrm ABSTRACT}
\vglue0.2cm
{\rightskip=3pc
 \leftskip=3pc
 \eightrm\baselineskip=10pt\noindent
We review some of the
recent work on the dynamics of four dimensional,
supersymmetric gauge theories.  The kinematics are largely determined
by holomorphy and the dynamics are governed by duality.  The results
shed light on the phases of gauge theories.  Some results and
interpretations which have not been published before
are also included.

\vglue0.6cm}

\tenrm\baselineskip=13pt
\newsec{Introduction}

Recently, it has become clear that certain aspects of four dimensional
supersymmetric field theories can be analyzed exactly, providing a
laboratory for the analysis of the dynamics of gauge theories (for a
recent elementary presentation and a list of references see
\ref\powerd{N. Seiberg, hep-th/9506077,
RU-95-37, IASSNS-HEP-95/46, to appear in the
Proc. of PASCOS 95, the Proc. of the Oskar Klein lectures, and in the
Proc. of the Yukawa International Seminar '95}).
For example, the phases of gauge theories and the mechanisms for phase
transitions can be explored in this context.  The dynamical mechanisms
explored are standard to gauge theories and thus, at
least at a qualitative level, the insights obtained
are expected to also be applicable for non-supersymmetric theories.
We summarize some of the
recent ideas.  The discussion is
not in historical order and other examples appear in the literature.

\subsec{Phases of gauge theories}

The phases of gauge theories can be characterized by the potential
$V(R)$ between electric test charges separated by a large
distance $R$.  Up to a non-universal, additive constant, the
potential is conjectured to behave as
\eqn\wlp{\eqalign{\hbox{Coulomb}\qquad V(R)& \sim {1\over
R}\cr \hbox{free electric}\qquad V(R)& \sim {1\over
R\log(R\Lambda)}\cr
\hbox{free magnetic}\qquad V(R) & \sim {\log(R\Lambda)\over R}\cr
\hbox{Higgs}\qquad V(R)& \sim constant \cr
\hbox{confining}\qquad V(R)& \sim  \sigma R.\cr}}
The first three phases have massless gauge fields and
potentials of the form $V(R)\sim e^2(R)/R$.  In the Coulomb
phase, the electric charge $e^2(R)\sim$constant.  In the free electric
phase, massless electrically charged fields renormalize the charge to
zero at long distances as $e^{-2}(R)\sim \log(R\Lambda )$.  Similar
behavior occurs when the long distance theory is a non-Abelian theory
which is not asymptotically free.  The free magnetic phase occurs when
there are massless magnetic monopoles, which renormalize the electric
coupling constant to infinity at large distance with a conjectured
behavior $e^2(R)\sim \log (R\Lambda )$.  In the Higgs
phase, the condensate of an electrically charged field gives a mass
gap to the gauge fields by the Anderson-Higgs mechanism and screens
electric charges, leading to a potential which, up to the additive
non-universal constant, has an exponential Yukawa decay to zero at
long distances.  In the confining phase, there is a mass gap
with electric flux confined into a thin tube, leading to the linear
potential with string tension $\sigma$.

All of the above phases can be non-Abelian as well as Abelian.  In
particular, in addition to the familiar Abelian Coulomb phase, there
are theories which have a non-Abelian Coulomb phase with massless
interacting quarks and gluons exhibiting the above Coulomb potential.
This phase occurs when there is a non-trivial, infrared fixed point of
the renormalization group.  These are thus non-trivial, interacting
four dimensional conformal field theories.

We can also consider the behavior of the potential $V(R)$ for
magnetic test charges separated by a large distance $R$.  Up to
an additive, non-universal constant, the potential behaves as
\eqn\tlp{\eqalign{\hbox{Coulomb}\qquad V(R)& \sim{1\over R}\cr
\hbox{free electric}\qquad V(R)&\sim {\log(R\Lambda)\over R}\cr
\hbox{free magnetic}\qquad V(R)& \sim{1\over R\log(R\Lambda)}\cr
\hbox{Higgs}\qquad V(R)& \sim \rho R \cr
\hbox{confining}\qquad V(R)& \sim constant. \cr}}
The behavior in the first three phases can be written as
$V(R)=g^2(R)/R$ where the effective
magnetic charge $g^2(R)$ is related to the
effective electric charge appearing in \wlp\ by the Dirac condition,
$e(R)g(R)\sim 1$.  The linear potential in the Higgs phase reflects
the string tension in the Meissner effect.

The above behavior is modified when there are matter fields in the
fundamental representation of the gauge group because virtual pairs
can be popped {}from the vacuum and completely screen the sources.
Indeed, in this situation there is no invariant distinction between
the Higgs and the confining phases \higgscon.  In particular, there is
no phase with a potential behaving as the ``confining'' potential in
\wlp\ at large distances -- the flux tube can break.
For large expectation values of the fields, a Higgs description is
most natural while, for small expectation values, it is more natural
to interpret the theory as ``confining.''  Because there is really no
distinction, it is possible to smoothly interpolate from one
interpretation to the other.

Note that under electric-magnetic duality, which exchanges
electrically charged fields with magnetically charged fields, the
behavior in the free electric phase is exchanged with that of the free
magnetic phase.  Mandelstam and 'tHooft suggested that, similarly, the
Higgs and confining phases are exchanged by duality.  Confinement can
then be understood as the dual Meissner effect associated with a
condensate of monopoles.  As we will review, in supersymmetric
theories it is possible to show that this picture is indeed correct.

Dualizing a theory in the Coulomb phase, we remain in the same phase
(the behavior of the potential is unchanged).  For an Abelian Coulomb
phase with free massless photons, this follows from a standard duality
transformation.  What is not obvious is that this is also the case in
a non-Abelian Coulomb phase.  This was first suggested by Montonen and
Olive
\ref\mo{C. Montonen and D. Olive, \pl {72}{1977}{117}; P. Goddard,
J. Nuyts and D. Olive, \np{125}{1977}{1}.}.
The simplest version of their proposal is
true only in $N=4$ supersymmetric field theories
\ref\dualnf{H. Osborn, \pl{83}{1979}{321}; A. Sen, hep-th/9402032,
\pl{329}{1994}{217}; C. Vafa and E. Witten, hep-th/9408074,
\np{432}{1994}{3}.}
\nref\swii{N. Seiberg and E. Witten, hep-th/9408099,
\np{431}{1994}{484}.}%
\nref\ntwomatti{A. Hanany and Y. Oz, TAUP-2248-95, WIS-95/19,
hep-th/9505075.}%
\nref\ntwomattii{P.C. Argyres, M.R. Plesser and A. Shapere,
IASSNS-HEP-95/32, UK-HEP/95-06, hep-th/9505100.}%
and in finite $N=2$
supersymmetric theories
\refs{\swii , \ntwomatti , \ntwomattii}.
The extension of these ideas to asymptotically free $N=1$ theories
first appeared in \sem\ and will be reviewed here.

\subsec{Super Yang-Mills theories}

We briefly review standard lore concerning $N=1$ supersymmetric
Yang-Mills theories -- i.e. pure super glue with no matter.  The lore
presented here can be proven via the analysis in the following
sections, by adding vector-like matter and then integrating it out.

We consider a theory based on a simple group $G$; the generalization
to semi-simple Yang-Mills theories is obvious.  The theory consists of
the $G$ vector bosons and gauginos $\lambda _{\alpha}$ in the adjoint
of $G$.  There is a classical $U(1)_R$ symmetry, gaugino number, which
is broken to a discrete $Z_{2h}$ subgroup by instantons, $\ev{(\lambda
\lambda )^h}={\rm const.}\Lambda ^{3h}$, where $h=C_2(A)$ is the
Casimir in the adjoint normalized so that, for example, $h=N_c$ for
$SU(N_c)$.  The lore is that this theory confines, gets a mass gap,
and that there are $h$ vacua associated with the spontaneous breaking
of the $Z_{2h}$ symmetry to $Z_2$ by gaugino condensation,
$\ev{\lambda \lambda}={\rm const.}\Lambda ^3$.  These vacua each
contribute $(-1)^F=1$ and thus the Witten index
\ref\wittentr{E. Witten,
\np{202}{1982}{253}} is $\Tr
(-1)^F=C_2(A)$.

\subsec{Outline}

In sect.\ 2, we discuss general techniques for analyzing
supersymmetric theories.  In sect.\ 3, we discuss the classical moduli
spaces of degenerate vacua which supersymmetric gauge theories with
matter often have.  In particular, we discuss the classical vacuum
degeneracy for $N=1$ supersymmetric QCD.  In sect.\ 4, we discuss
supersymmetric QCD for $N_f\leq N_c+1$ massless flavors of quarks in
the fundamental representation of $SU(N_c)$. In sect.\ 5, we discuss
$N_f>N_c+1$ and duality.  Sect.\ 6 is a brief discussion of the phases
and duality of supersymmetric $SO(N_c)$ with matter fields in the
$N_c$ dimensional representation.  In sect.\ 7, we discuss general
aspects of supersymmetric theories which have a low energy Abelian
Coulomb phase.  In sect.\ 8, we consider the examples of $SU(2)$ with
$N_f=1$ and $N_f=2$ adjoints, which have electric-magnetic-dyonic
triality.

\newsec{Holomorphy and symmetries}

\subsec{General considerations}

The basic approach will be to consider the low energy effective action
for the light fields, integrating out degrees of freedom above some
scale.  Assuming that we are working above the scale of possible
supersymmetry breaking, the effective action will have a linearly
realized supersymmetry which can be made manifest by working in terms
of superfields.  The light matter fields can be combined into chiral
superfields $X_r=\phi _r+\theta _\alpha \psi ^\alpha _r+\dots$, where
the $\phi _r$ are scalars and the $\psi ^\alpha _r$ are Weyl fermions.
In addition, there are the conjugate anti-chiral superfields
$X_r^\dagger =\phi _r^\dagger +\bar \theta _{\dot\alpha}\psi ^{\dagger
\dot\alpha}_r+\dots$.   Similarly, light gauge fields
combine into supermultiplets involving a gauge boson $A_\mu$
and gauginos $\lambda _\alpha$ and $\lambda ^\dagger _{\dot \alpha}$.

We will focus on a particular contribution to the effective Lagrangian
-- the superpotential term
\eqn\supi{\int d^2\theta W_{\rm eff}(X_r,g_I, \Lambda),}
where $X_r$ are the light fields, the $g_I$ are various coupling
constants, and $\Lambda$ is the dimensional transmutation scale
associated with the gauge dynamics,  $-{8\pi^2\over g^2(\mu)}\sim \log
\Lambda/\mu$.
Upon doing the $\theta$ integral, the superpotential yields a potential
for the scalars and a Yukawa type interaction with the scalars and the
fermions.

The key fact is that supersymmetry requires $W_{\rm eff}$ to be
holomorphic in the chiral superfields $X_r$, independent of the
$X_r^{\dagger}$.

We will think of all the coupling constants $g_I$ in the tree level
superpotential $W_{\rm tree}$ and the scale $\Lambda$
as background fields \nonren.  Then, the
quantum, effective superpotential, $W_{\rm eff}(X_r, g_I,
\Lambda)$ is constrained by:

\lfm{1.} Symmetries and selection rules:
By assigning transformations laws both to the
fields and to the coupling constants, the theory has a large symmetry.
The effective Lagrangian should be invariant under it.

\lfm{2.} Holomorphy: $W_{\rm eff}$ is independent of
$g_I^\dagger$ \nonren.
This is the key property.  Just as the superpotential is holomorphic in
the fields, it is also holomorphic in the coupling constants (the
background fields).  This is unlike the effective Lagrangian in
non-supersymmetric theories, which is not subject to any holomorphy
restrictions.  This use of holomorphy extends considerations of
\lref\polsei{J. Polchinski and N. Seiberg, (1988) unpublished.}
\refs{\cern, \sv, \polsei}. It is similar in spirit to the proofs of
non-renormalization in sigma model perturbation theory
\ref\witnonren{E. Witten, \np{268}{1986}{79}} and in
semi-classical perturbation theory
\ref\dsnonren{M. Dine and N. Seiberg, \prl{57}{1986}{2625}
} in string theory.

\lfm{3.} Various limits: $W_{\rm eff}$ can be analyzed approximately at
weak coupling.  The singularities have physical meaning and can be
controlled.

\medskip

Often these conditions completely determine $W_{\rm eff}$.  The point
is that a holomorphic function (more precisely, a section) is
determined by its asymptotic behavior and singularities.  The results
can be highly non-trivial, revealing interesting non-perturbative
dynamics.

When there is a Coulomb phase, the kinetic terms for the gauge fields
are also constrained
by the above considerations.  The relevant term in the effective
Lagrangian is
\eqn\gkti{\int d^2\theta\ Im [\tau _{\rm eff}(X_r,g_I, \Lambda
)W_\alpha ^2];} $W_{\alpha}^2$ gives the supersymmetric completion of
$F^2+iF\widetilde F$ so
\eqn\tauis{\tau _{\rm eff}\sim {\theta _{\rm
eff}\over 2\pi}+{4\pi i\over g_{\rm eff}^2}}
is the effective gauge coupling constant.  $\tau _{\rm eff}(X_r,g_I,
\Lambda )$ is
holomorphic in its arguments and can often be exactly determined.

\subsec{Example: Wess-Zumino Model}

In order to demonstrate the above rules,
we consider the simplest Wess-Zumino
model and rederive the known non-renormalization theorem.  Consider the
theory based on the tree level superpotential
\eqn\wztr{ W_{\rm tree} = m \phi^2 + g\phi^3 . }
We will make use of two $U(1)$ symmetries.  The charges of the field
$\phi$ and the coupling constants $m$ and $g$ are
\eqn\wzch{\matrix{\quad& U(1)& \times & U(1)_R \cr
\phi & 1 && 1 \cr
m & - 2 && 0 \cr g & -3 && -1 \cr}} where $U(1)_R$ is an $R$ symmetry
under which $d^2\theta $ has charge $-2$ and thus, in order for \supi\
to be invariant, the superpotential has charge 2.  Note that non-zero
values for $m$ and $g$ explicitly break both $U(1)$ symmetries.
However, they still lead to selection rules.

The symmetries and holomorphy of the effective superpotential restrict
it to be of the form
\eqn\wzwsym{ W_{\rm eff}= m \phi^2 f\left( {g \phi \over m} \right). }
Consider the limit $g\rightarrow 0$ and $m\rightarrow 0$ with
arbitrary $g/m$.  We must
have $W_{eff}\rightarrow W_{tree}$ and, therefore,
$f(t=(g\phi/m))\rightarrow 1+t$ in this limit of vanishing coupling.
Now, because $t$ is arbitrary in this limit, $f(t)$ is thus evaluated
to be given by $f(t)=1+t$ for all $t$.  Therefore, the exact effective
superpotential is found to be
\eqn\wznoren{W_{\rm eff}= m \phi^2 + g \phi^3 = W_{\rm tree};}
i.e.\ the superpotential is not renormalized \nonren.

This argument rederives the standard perturbative
non-renormalization theorem \ref\pertnonren{
M.T. Grisaru, W. Siegel and M. Rocek,
\np{159}{1979}{429}} and
extends it beyond perturbation theory.  Strictly
speaking, the Wess-Zumino model probably does not exist as an
interacting quantum field theory in four dimensions except as an
effective low energy theory of a more fundamental theory
and, therefore, this non-perturbative result is only of limited
interest.  This non-perturbative proof of the Wess-Zumino
non-renormalization can be directly applied to
two dimensions, where such models do exist as interacting quantum
field theories.

If there are several fields, some light and some heavy, the heavy
fields can be integrated out to yield a low energy effective
Lagrangian for the light fields.  The contribution of tree diagrams
with intermediate heavy fields should be then included in the
effective action.  The above simple rules allow such diagrams to
contribute and are thus compatible with the known tree level
renormalization of the superpotential.

\subsec{The 1PI effective action}

There are two different objects which are usually called ``the
effective action:'' the 1PI effective action and the Wilsonian one.
When there are no interacting massless particles, these two effective
actions are identical.  This is often the case in the Higgs or
confining phases.  However, when interacting massless particles are
present, the 1PI effective action suffers from IR ambiguities and
might suffer from holomorphic anomalies
\ref\sv{M.A. Shifman and A. I. Vainshtein, \np{277}{1986}{456};
\np{359}{1991}{571}.}.
These are absent in the Wilsonian effective action.

Consider the theory with a tree level superpotential with sources for
the gauge invariant polynomials $X^r$ in the matter fields, $W_{\rm
tree}=\sum _r g_rX^r$, with the $g_r$ regarded as background chiral
superfield sources \nonren.  The functional integral with the added
source terms gives the standard generating function for the
correlation functions, $\Gamma(g)$.  If supersymmetry is not broken,
$\Gamma(g)$ is supersymmetric (otherwise we should include the
Goldstino field and supersymmetry will be realized non-linearly)
and\foot{In writing this expression we should think of the coupling
constants $g_r$ as background superfields.  Otherwise, $W_L(g)$ is a
constant superpotential, which has no effect in global supersymmetry.
Indeed, the following equation can be interpreted as differentiating
the action with respect to the $F$ component of $g_r$.}  $\Gamma
(g)=\dots +\int d^2\theta W_L(g)$.  Using $W_L(g)$ we can compute the
expectation values
\eqn\corr{{\partial W_L(g)\over \partial g_r}=\ev{X^r}.}
It is standard to perform a Legendre transform to find the 1PI
effective action for the operators $X_r$:
\eqn\invlt{W_{\rm dyn}(X^r)=\left(W_L(g_r)-\sum _rg_rX^r\right)_{\ev
{g_r}},} where the $\ev{g_r}$ are the solutions of \corr.  The
transformation {}from $W_L(g_r)$ to $W_{\rm dyn}(X_r)$ can be inverted
by the inverse Legendre transform as
\eqn\wl{W_L(g)=\left(W_{\rm dyn}(X^r)+
\sum _rg_rX^r\right) _{\ev{X^r}},}
where the $X^r$ are evaluated at their expectation values $\ev{X^r}$,
which solve
\eqn\eomg{{\partial W_{\rm dyn}\over \partial
X_r}+g_r=0.}

The 1PI effective superpotential
\eqn\weffopi{W_{\rm eff}(X,g)=W_{\rm dyn}(X^r)+\sum _rg_rX^r }
has the property that the equations of motion \eomg\ for the fields
$X^r$ derived from it determine their expectation values.  In some
cases the superpotential $W_{\rm eff}$ obtained by the above Legendre
transform is the same as the Wilsonian superpotential for the light
fields.  In applying this procedure we should be careful of the
following pitfalls:

\item{1.} The theory with the
sources should have a gap.  Otherwise, the
1PI action is ill defined.

\item{2.}  The theory with the sources might break supersymmetry.
In that case $W_L$ is ill defined.

\item{3.} As the sources are turned off, some particles
become massless.
Their interpolating fields should be among the composite fields $X^r$.
If some massless particles cannot be represented by a gauge invariant
operator $X^r$, the effective superpotential derived this way will not
include them.  This often leads to singularities.

\item{4.} The theory might also have other branches
which are present only when some sources vanish.  In this case
there are new massless particles at that point and the 1PI action might
miss some of the branches.  In other words, then the Legendre transform
does not exist.

\item{5.} If some composites do not represent massless particles, they
should be integrated out.  Although we can use the effective
superpotential to
find their expectation values, we cannot think of them
as fields corresponding to massive particles except near a point where
they become massless.

\noindent
There are known
examples\foot{As a  simple example of a
situation in which the Legendre transform analysis is incomplete
because supersymmetry is dynamically broken by the added source terms,
consider $N=1$ supersymmetric $SU(2)$ with a single field
$Q$ in the $\bf{4}$ of $SU(2)$
\ref\iss{K. Intriligator, N. Seiberg, and S. Shenker,
\pl{342}{1995}{152}.}.
The theory without added source terms has a one complex dimensional
smooth moduli space of vacua labeled by $\ev{X}$, where $X=Q^4$ is the
basic gauge invariant, with a superpotential $W(X)=0$.
Adding a source
$W=gX$ does not lead to a
supersymmetric effective superpotential $W(g)$
-- rather, it breaks supersymmetry \iss.  (As discussed in \iss, it is
also possible
that there is a non-Abelian Coulomb phase at the origin of
the moduli space and that supersymmetry is unbroken with the added
source term.  In that case the 1PI analysis again fails to capture the
physics.)} of each of these situations; in these cases the
1PI effective action
is misleading, failing to capture important aspects
of the physics.

When we can use this procedure to find the Wilsonian action, the
linearity of $W_{\rm
eff}$ \weffopi\ in the sources provides a derivation of
the linearity of the Wilsonian effective action in the sources.
\lref\iipi{C.P. Burgess, J.-P. Derendinger, F. Quevedo, and M. Quiros,
hep-th/9595171, CERN-Th/95-111.}
See \refs{\kaplou, \iipi} for a related discussion.

This approach is particularly useful when we know how to compute
$W_L(g_r)$ exactly.  Then, $W_{\rm dyn}$ and $W_{\rm eff}$ follow
simply from the Legendre transform \invlt; this is the ``integrating
in'' discussed in \ilsii .  One situation where $W_L(g_r)$ can be
determined is when the $X^r$ are all quadratic in the elementary
fields.  In that case, the sources $g_r$ are simply mass terms for the
matter fields and $W_L(g)$ is the superpotential for the low energy
gauge theory with the massive matter integrated out, expressed in
terms of the quantities in the high-energy theory.

These issues will be exhibited and
further discussed in later sections.

\subsec{Extended supersymmetry}

Theories with extended supersymmetry are further constrained.  For
example, $N=2$ supersymmetry combines an $N=1$ vector superfield with
a chiral superfield $\Phi$ in the adjoint representation of the gauge
group into an $N=2$ vector multiplet.  In particular, $N=2$
supersymmetry includes a global $SU(2)_R$ symmetry which rotates the
gluino component of an $N=1$ vector superfield into the fermion
component of the chiral superfield.  This symmetry relates the Kahler
potential for the chiral superfield to the $\tau _{eff}$ in the
kinetic term \gkti\ for the gauge field.  Therefore, the Kahler
potential for chiral superfields which are part of $N=2$ vector
multiplets are determined in terms of holomorphic functions and can thus often
be obtained exactly.

Another type of $N=2$ supermultiplet is the hypermultiplet, consisting
of two $N=1$ chiral multiplets $Q$ and $\widetilde Q$.  The global
$SU(2)_R$ symmetry rotates the scalar component of $Q$ into the scalar
component of $\widetilde Q^\dagger$.  This symmetry implies that the
Kahler potential for these fields must yield a metric which is
hyper-Kahler.  $N=2$ supersymmetry further implies that this metric is
not corrected by quantum effects.

\nref\bogr{E. B. Bogomol'nyi,
Sov. J. Nucl.  Phys. {\bf 24} (1976) 449}\nref\psref{M. K. Prasad and
C. M.  Somerfeld, \prl{35}{1975}{760}}%
Another condition on theories with $N=2$ supersymmetry is the
Bogomol'nyi- Prasad, Sommerfeld bound
\refs{\bogr, \psref}, which is related to
the central term in the $N=2$ algebra \ref\wor{E. Witten and D. Olive,
\pl{78}{1978}{97}}.  The mass of any state satisfies
\eqn\bpsbnd{M\geq \sqrt{2}|Z|,}
where $Z$ is the central term of the $N=2$ algebra, involving the
gauge and global quantum numbers of the state.  The BPS saturated
states, for which the inequality \bpsbnd\ is saturated, are some of
the stable states in the spectrum.  Because $Z$ is a holomorphic
object, it can often be exactly determined; \bpsbnd\ then yields the
exact mass spectrum for some of the stable states.

This review will focus on $N=1$ supersymmetry.  We refer the
reader to the literature for further details about extended
supersymmetry.

\newsec{Classical super gauge theories --
classical moduli spaces of vacua}

The classical Lagrangian of a supersymmetric theory with gauge group
$G$, matter superfields $\Phi _f$ in representations $R(f)$ of $G$,
and zero tree level superpotential is
\eqn\clglag{\lL=\lL _{o}+\sum _{f,a}\phi _f^\dagger \lambda ^aT^a_f\psi
_f+h.c. +\sum _a(\sum _f\phi _f^\dagger T^a_f\phi _f)^2,}
where $\lL _{o}$ are the obvious gauge invariant kinetic terms for the
gauge and matter fields and $T^a_f$
are the $G$ generators in representation $R(f)$.  The interactions in
\clglag\ are related by supersymmetry to
the coupling in $\lL _o$ of gauge fields to matter.

Classical gauge theories often have ``$D$-flat'' directions of
non-zero $\ev{\phi _f}$ along which the squark potential in \clglag\
vanishes.  In other words, these theories often have classical moduli
spaces of degenerate vacua.  As a simple example, consider $U(1)$
gauge theory with a matter superfield $Q$ of charge 1 and $\widetilde
Q$ of charge $-1$.  The squark potential in \clglag\ is $V=(Q^\dagger
Q-\widetilde Q^\dagger \widetilde Q)^2$ and thus there is a continuum
of degenerate vacua labeled, up to gauge equivalence, by
$\ev{Q}=\ev{\widetilde Q}=a$, for any complex $a$.  In vacua with
$a\neq 0$ the gauge group is broken by the super Higgs mechanism.  The
gauge superfield gets mass $|a|$ by ``eating'' one chiral superfield
degree of freedom from the matter fields.  Since we started with the
two superfields $Q$ and $\widetilde Q$, one superfield degree of
freedom remains massless.  The massless superfield can be given a
gauge invariant description as $X=Q\widetilde Q$.  In the vacuum
labeled as above by $a$, $\ev{X}=a^2$.  Because $a$ is arbitrary,
there is no potential for $X$, $W_{cl}(X)=0$ -- classically $X$ is a
``modulus'' field whose expectation value labels a classical moduli
space of degenerate vacua.  The classical Kahler potential of the
microscopic theory is $K_{cl}=Q^\dagger e^{V}Q+\widetilde Q^\dagger
e^{-V}\widetilde Q$.  In terms of the light field $X$,
$K_{cl}=2\sqrt{X^\dagger X}$, which has a conical ($Z_2$ orbifold)
singularity at $X=0$.  A singularity in a low energy effective action
generally reflects the presence of additional massless fields which
should be included in the effective action.  Indeed, the singularity
at $X=0$ corresponds to the fact that the gauge group is unbroken and
all of the original microscopic fields are classically massless there.

As in the above simple example, the classical moduli space of vacua
is the space of squark expectation values $\ev{\phi
_f}$, modulo gauge equivalence, along which the potential in \clglag\
vanishes.
It can always be given a gauge invariant description in terms
of the space of expectation values $\ev{X_r}$ of gauge invariant
polynomials in the fields subject to any classical relations.
This is because
setting the potential in \clglag\ to zero and modding out
by the gauge group is equivalent to modding out by the complexified
gauge group.   The space of chiral superfields modulo the complexified
gauge group can be parameterized by the gauge invariant polynomials
modulo any classical relations. These results follow from results in
geometrical invariant theory
\ref\mumford{D. Mumford and J. Fogarty,
{\it Geometric Invariant Theory} (Springer, 1982).}; see also
\ref\luttay{M. A. Luty and W. Taylor IV, hep-th/9506098, MIT-CTP-2440
} for a recent discussion.  As in the above example, the fields $X_r$
correspond to the matter fields left massless after the Higgs
mechanism and are classical moduli, $W_{cl}(X_r)=0$.

The vacuum degeneracy of classical moduli spaces of vacua is not
protected by any symmetry.  In fact, vacua with different expectation
values of the fields are physically inequivalent: as in the above
example, the masses of the massive vector bosons depend on the
expectation values $\ev{X_r}$.  Therefore, the degeneracy of a
classical moduli space of vacua is accidental and can be lifted in the
quantum theory by a dynamically generated $W_{\rm eff}(X_r)$.  We will
often be able to determine $W_{\rm eff}(X_r)$ by the considerations
discussed above.
\vglue0.3cm
\leftline{\it Example: SUSY QCD}
\vglue0.3cm
Consider supersymmetric $SU(N_c)$ gauge theory with $N_f$ quark
flavors $Q^i$ in the fundamental representation and $\widetilde
Q_{\tilde i}$ in the anti-fundamental, $i, \tilde i=1\dots N_f$.  In
the absence of mass terms, there is a classical moduli space of vacua
given, up to gauge and global symmetry transformations, by:
\eqn\flatqi{ Q=\widetilde Q= \pmatrix { a_1& & & & \cr
&a_2& & & \cr
& & .
& & \cr
& & & a_{N_f} &\quad \cr}}
for $N_f < N_c$, with $a_i$ arbitrary, and by
\eqn\flatqii{Q=
\pmatrix { a_1& & & \cr
&a_2& & \cr
& & .  & \cr
& & & a_{N_c}\cr
& & & \cr
& & & \cr} , \quad
\widetilde Q= \pmatrix {
\widetilde a_1& &       &         \cr
   &\widetilde a_2&     &         \cr
   &   & .   &         \cr
   &   &     & \widetilde a_{N_c}\cr
   &   &     &         \cr
   &   &     &         \cr}, }
$$ |a_i|^2 - |\widetilde a_i|^2= {\rm independent ~ of} ~ i, $$
for $N_f \ge N_c$.

For $N_f<N_c$, the gauge invariant description of the classical moduli
space is in terms of arbitrary expectation values of the ``mesons''
$M^i_{\tilde j}=Q^i\widetilde Q_{\tilde j}$.  For $N_f\geq N_c$, it is
also possible to form ``baryons'' $B^{i_1\dots i_{N_c}}=Q^{i_1}\dots
Q^{i_{N_c}}$ and $\widetilde B _{\tilde i_1\dots \tilde
i_{N_c}}=\widetilde Q_{\tilde i_1}\dots \widetilde Q_{\tilde
i_{N_c}}$.  The gauge invariant description of the classical moduli
space for $N_f\geq N_c$ is given in terms of the expectation values of
$M$, $B$ and $\widetilde B$, subject to the following classical
constraints.  Up to global symmetry transformations, the expectation
values are
\eqn\flatmbtb{\eqalign{
M&=\pmatrix { a_1 \widetilde a_1& & &  & & &\cr
&a_2 \widetilde a_2 & &  & & &\cr
& & .  & & & & \cr
& & & a_{N_c}\widetilde a_{N_c} & & &\cr
& & & & & & \cr
& & & & & & \cr} \cr
B^{1,...,N_c}&=a_1a_2...a_{N_c} \cr
\widetilde B_{1,...,N_c}&=
\widetilde a_1 \widetilde  a_2...\widetilde a_{N_c} \cr}}
with all other components of $M$, $B$ and $\widetilde B$ vanishing.
Therefore, the rank of $M$ is at most $N_c$.  If it is less than
$N_c$, either $B=0$ with $\widetilde B$ having rank at most one or
$\widetilde B=0$ with $B$ having rank at most one.  If the rank of $M$
is equal to $N_c$, both $B$ and $\widetilde B$ have rank one and the
product of their eigenvalues is the same as the product of non-zero
eigenvalues of $M$.

As discussed above, the physical interpretation of the flat directions
is that the gauge group is Higgsed.  If $B=\widetilde B=0$ and $M$ has
rank $k$, $SU(N_c)$ is broken to $SU(N_c-k)$ with $N_f-k$ light
flavors.

Already at the classical level, we can integrate out the massive
fields and consider an effective Lagrangian for the massless modes.
Their expectation values label the particular ground state we expand
around, and hence they are coordinates on the classical moduli space.
The classical moduli space is not smooth.  Its singularities are at
the points of enhanced gauge symmetry.  For instance, when
$a_i=\widetilde a_i=0$ for every $i$ the gauge symmetry is totally
unbroken.  Therefore, the low energy effective theory of the moduli is
singular there.  This should not surprise us.  At these singular
points there are new massless particles -- gluons.  An effective
Lagrangian without them is singular.  If we include them in the low
energy description, the Lagrangian is smooth.

In the next two sections we will see how this picture changes in the
quantum theory.  At large expectation values of the fields, far from
the classical singularities, the gauge symmetry is broken at a high
scale, the quantum theory is weakly coupled, and semi-classical
techniques are reliable.  We expect the quantum corrections to the
classical picture to be small there.  On the other hand, at small
field strength the quantum theory is strongly coupled and the quantum
corrections can be large and dramatically modify the classical
behavior.  In particular, the nature of the classical singularities,
which are at strong coupling, is generally totally different in the
quantum theory.

\newsec{SUSY QCD for $N_f\leq N_c+1$}

Because these theories have matter fields in the fundamental
representation of the gauge group, as mentioned in the introduction,
there is no invariant
distinction between the Higgs and the confining phases \higgscon.
It is possible
to smoothly interpolate from one interpretation to the other.

\subsec{$N_f < N_c$ -- No Vacuum}

The first question to ask is whether the classical vacuum degeneracy
can be lifted quantum mechanically by a dynamically generated
superpotential. The form of such a superpotential is constrained by
the symmetries.  At the classical level, the symmetries are
\eqn\globsymc{SU(N_f)_L\times SU(N_f)_R
 \times U(1)_A\times U(1)_B \times U(1)_R }
where the the quarks transform as
\eqn\qtransl{\eqalign{
Q &\qquad (N_f,1,1,1,{N_f-N_c \over N_f}) \cr
\widetilde Q & \qquad
(1, \overline N_f,1,-1,{N_f-N_c \over N_f}). \cr}}
$U(1)_R$ is an $R$ symmetry (the gauginos have charge $+1$, the squark
components of $Q$ and $\widetilde Q$ have the charge $R(Q)$ indicated
above, and the charge of the fermion components is
$R(\psi)=R(Q)-1$). The charges were chosen so that only the symmetry
$U(1)_A$ is anomalous in the quantum theory.  Considering the
anomalous $U(1)_A$ as explicitly broken by fermion zero modes in an
instanton background, $U(1)_A$ leads to a selection rule.  The
instanton amplitude is proportional to $e^{-S_{Inst.}}=e^{-8\pi
^2g^{-2}(\mu)+i\theta }=(\Lambda/\mu) ^{3N_c-N_f}$, where $\Lambda$ is
the dynamically generated scale of the theory, and we integrated the
1-loop beta function.  Therefore, $U(1)_A$ is respected provided we
assign $\Lambda ^{3N_c-N_f}$ charge $2N_f$ to account for the charge
of the fermion zero modes under $U(1)_A$.  The dependence on the scale
$\Lambda$ is thus determined by the $U(1)_A$ selection rule.

There is a unique superpotential
which is compatible with these symmetries
\ref\dds{A.C. Davis, M. Dine and N. Seiberg,
\pl{125}{1983}{487}}
\eqn\wsyma{W_{\rm eff} =
C_{N_c,N_f}\left({\Lambda^{3N_c-N_f} \over \det
\widetilde QQ}\right)^{1/(N_c-N_f)},}
where $C_{N_c,N_f}$ are constants which depend on the subtraction
scheme.  Therefore, if the vacuum degeneracy is lifted, this
particular superpotential must be generated.  For $N_f \ge N_c$ this
superpotential does not exist (either the exponent diverges or the
determinant vanishes) and therefore the vacuum degeneracy cannot be
lifted.  We will return to $N_f\geq N_c$ in the next subsections.

Note that the superpotential \wsyma\ is non-perturbative and is thus
not in conflict with the {\it perturbative} non-renormalization
theorem.  Indeed, the above argument demonstrates that the
perturbative non-renormalization and its non-perturbative violation
can be understood simply as the need for obtaining a well-defined
charge violation of the anomalous $U(1)_A$ as occurs, for example, in
an instanton background.

The superpotential \wsyma\ is further constrained by considering
various limits. For example, consider the limit of large
$M^{N_f}_{\tilde N_f}$, i.e. large $a_{N_f}$ in \flatqi, which breaks
$SU(N_c)$ with $N_f$ flavors to $SU(N_c-1)$ with $N_f-1$ light flavors
by the Higgs mechanism at energy $a_{N_f}$.  Matching the running
gauge coupling at energy $a_{N_f}$, the low energy theory has scale
$\Lambda _L^{3(N_c-1)-(N_f-1)}=\Lambda ^{3N_c-N_f}/a_{N_f}^2$.  The
fact that the scales are so matched without any threshold factors
reflects a choice of subtraction scheme; this is the correct
matching, for example, in the $\overline{DR}$ scheme.  Requiring
\wsyma\ to properly reproduce the superpotential of the low energy
theory in this limit gives $C_{N_c,N_f}=C_{N_c-N_f}$.

Next consider giving $Q_{N_f}$ and $\widetilde Q_{\tilde N_f}$ a large
mass by adding $W_{\rm tree}=mM^{N_f}_{\tilde N_f}$.  The low energy
theory is $SU(N_c)$ with $N_f-1$ flavors.  Matching the running gauge
coupling at the transition scale $m$, the low energy theory has scale
$\Lambda _L^{3N_c-(N_f-1)}=m\Lambda ^{3N_c-N_f}$.  Again, this
equality is up to a scheme dependent threshold factor which is one in
the $\overline{DR}$ scheme.  Using the symmetries, the exact
superpotential with the added mass term is of the form
\eqn\wexactf{W_{\rm exact}=\left({\Lambda ^{3N_c-N_f}\over
\det M}\right)^{1/(N_c-N_f)}f\left(t=mM^{N_f}_{\tilde N_f}
\left({\Lambda ^{3N_c-N_f}\over \det M}\right)^{-1/(N_c-N_f)}\right).}
In the limit of small mass and weak coupling, we know that
$f(t)=C_{N_c,N_f}+t$.  Because all values of $t$ can be obtained in
this limit, the function $f(t)$ is evaluated in this understood limit
to be $f(t)=C_{N_c,N_f}+t$ for all $t$.  The exact superpotential with
the added mass term is thus
\eqn\wexactnof{W_{\rm exact}=C_{N_c,N_f}\left({\Lambda ^{3N_c-N_f}\over
\det M}\right)^{1/(N_c-N_f)}+mM^{N_f}_{\tilde N_f}.}
Requiring \wexactnof\ to
give the correct superpotential in the low energy theory upon
integrating out $M^{N_f}_{\tilde N_f}$ relates $C_{N_c,N_f}$ to
$C_{N_c,N_f-1}$ which, when combined with $C_{N_c,N_f}=C_{N_c-N_f}$,
determine that $C_{N_c,N_f}=(N_c-N_f)C^{1/(N_c-N_f)}$ with $C$ a
universal constant.

For $N_f=N_c-1$, the superpotential \wsyma\ is proportional to the
one-instanton action and, thus, the constant $C$ can be exactly
computed via a one-instanton calculation.  Because the gauge group is
completely broken by the Higgs mechanism for $N_f=N_c-1$ (for $\det M
\neq 0$), the instanton calculation is reliable (there is no
infra-red divergence).  The universal constant $C$ can be computed by
considering the particular case $N_c=2$, $N_f=1$.  The direct
instanton calculation \ads\ reveals that the constant $C\neq 0$.  The
more detailed analysis of \finnpou\
shows that $C=1$ in the $\overline{DR}$
scheme.  For $N_f<N_c$ there is thus a dynamically generated
superpotential
\eqn\wsymi{W_{\rm eff} = (N_c-N_f)\left({\Lambda^{3N_c-N_f} \over \det
\widetilde Q Q }\right)^{1/(N_c-N_f)}.}

While $W_{\rm eff}$ is generated by instantons for $N_f=N_c-1$,
for $N_f < N_c-1$  it is associated with
gaugino condensation in the unbroken $SU(N_c-N_f)$ gauge group \ads.
In particular, the low energy theory has a WZ term\foot{We absorb a
factor of $1/32\pi ^2$ into the definition of $W_{\alpha}^2$}
\eqn\sunwzterm{\int d^2\theta \log (\det M)(W_\alpha W^\alpha)
_{SU(N_c-N_f)} =(M^{-1})_i^{\tilde j}F_{M^i_{\tilde j}}\lambda
_{\alpha}\lambda ^\alpha +\dots ,} to account for the matching to the
high energy theory.  If gaugino condensation occurs in the low energy
theory, $\ev {\lambda
\lambda}=c\Lambda _{N_c-N_f}^3$ where $c$ is a constant and $\Lambda
_{N_c-N_f}$ is the scale of the low energy $SU(N_c-N_f)$ gauge theory,
given in terms of the scale $\Lambda$ of the high-energy theory by
$\Lambda _{N_c-N_f}^{3(N_c-N_f)}=\Lambda ^{3N_c-N_f}/\det M$,
\sunwzterm\ yields a term $c\Lambda _{N_c-N_f}^3M^{-1}F_{M}$ in
the effective Lagrangian.  Our superpotential \wsymi\ indeed gives
exactly such a term, $(\Lambda ^{3N_c-N_f}/\det
M)^{1/(N_c-N_f)}M^{-1}F_M=
\Lambda _{N_c-N_f}^3M^{-1}F_M$, in the effective
Lagrangian.  Therefore, gaugino condensation occurs in $N=1$
$SU(N_c-N_f)$ Yang-Mills theory with the normalization
\eqn\gcsun{\ev{\lambda _\alpha \lambda ^\alpha}=e^{2\pi ik/(N_c-N_f)}
\Lambda _{N_c-N_f}^3 \qquad\hbox{with}\qquad k=1\dots (N_c-N_f),}
where we explicitly exhibit the phase.  Using
\wsymi (whose normalization follows from a well understood instanton
calculation), we have {\it derived} gaugino condensation \gcsun,
including its normalization in the $\overline{DR}$ scheme, in the low
energy $N=1$ Yang-Mills theory.  (For related work on this model see
also references \cern\ and
\ref\nsvz{V.A. Novikov, M.A. Shifman, A.I.  Vainshtein and V.I.
Zakharov, \np{260}{1985}{157}; A.I. Vainshtein, V.I. Zakharov, and
M.A. Shifman, Sov. Phys. Usp. {\bf 28} (1985) 709.}.)

The dynamically generated superpotential \wsymi\ leads
to a squark potential which
slopes to zero for $\det M\rightarrow \infty$.  Therefore, the quantum
theory does not have a ground state.
We started with an infinite set of
vacua in the classical theory and
ended up in the quantum theory without
a vacuum!

Consider adding $W_{\rm tree}=\Tr mM$, giving masses to the $N_f$
flavors.  As in \wexactnof, symmetries and the weak coupling and small
mass limit determine that the exact superpotential is $W_{\rm
full}=W_{\rm eff}+W_{\rm tree}$.  The vacua are given by $\ev{M}$
solving ${\partial W_{\rm full}\over
\partial M^{i}_{\tilde j}}|_{\ev{M}}=0$.  This gives $N_c$ vacua
\eqn\mevm{\ev{M^{i}_{\tilde j}}=(\det m\Lambda
^{3N_c-N_f})^{1/N_c}\left({1\over m}\right)^{i}_{\tilde j},}
corresponding to the $N_c$ branches of the $N_c$-th root.  For large
$m$, the matter fields are very massive and decouple, leaving a
low energy $SU(N_c)$ pure Yang-Mills theory.  Indeed, the low energy
theory has confinement with a mass gap and $N_c$ vacua.

Evaluating
\wsymi\ with the expectation values \mevm\ yields
\eqn\ymw{W_L(m)=N_c(\Lambda ^{3N_c-N_f}
\det m)^{1/N_c}=N_c\Lambda _L^3,}
where $\Lambda _L$ is the scale of the low energy $SU(N_c)$ Yang-
Mills theory.  Expressed in terms of $\Lambda _L$, the superpotential
\ymw\ is interpreted as the result of gaugino condensation in the low
energy $SU(N_c)$ Yang-Mills theory.  Expressed in terms of $m$ and
$\Lambda$, the superpotential \ymw\ is the effective superpotential
which yields the expectation values \mevm\ via \corr, ${\partial
W_L(m)\over \partial m}=\ev{M}$.  As in \invlt,
$W_L(m)$ leads to an effective superpotential for the operators $M$:
\eqn\sqcd{W_{\rm dyn}=(N_c-N_f)\left({\Lambda ^{3N_c-N_f}\over \det
M}\right)^{1/(N_c-N_f)}.}
In this case $W_{\rm dyn}$ agrees with the Wilsonian effective
superpotential \wsymi, as could have been expected because this
theory satisfies all of the conditions spelled out in sect.\ 2.3.
The fact that the Wilsonian effective action is here the same as
the 1PI effective action provides a simple alternate derivation
of the linearity derived in \wexactnof.

It is also possible to ``integrate in'' operators which do not
correspond to massless particles.  Then, the effective action can be
used only to compute their expectation values, rather than for
studying them as massive particles.  An example is the ``glueball''
field $S=-W_\alpha ^2$, whose source is $\log \Lambda ^{3N_c-N_f}$.
Integrating in $S$ by the Legendre transform of \sqcd\ with respect to
the source $\log \Lambda ^{3N_c-N_f}$ yields
\eqn\sqcds{W(S,M)=S\left[\log\left({\Lambda ^{3N_c-N_f}\over
S^{N_c-N_f}\det M}\right)+(N_c-N_f)\right],}
the superpotential obtained in \vy.
Adding mass terms $W_{tree}=\Tr mM$ and integrating out $M$ yields
\eqn\wsym{W(S)=S\left[\log\left({\Lambda _L^{3N_c}\over
S^{N_c}}\right)+N_c\right],} where $\Lambda _L$ is the scale of the
low energy $SU(N_c)$ Yang-Mills theory, $\Lambda _L^{3N_c}=\det
m\Lambda ^{3N_c-N_f}$.  This superpotential simply gives the
information discussed above: supersymmetric $SU(N_c)$ Yang-Mills
theory has $N_c$ supersymmetric vacua with the gaugino condensates
\gcsun\ and superpotential $W=N_c\Lambda _L^3$.  Working with such
effective superpotentials which include massive fields can be
convenient when interesting but complicated dynamics is encoded in the
integrating out of these massive fields.  Several such examples can be
found in \ils.  However, as stressed above, we should not think of $S$
as a field describing a massive particle.

\subsec{$N_f = N_c$ -- Quantum moduli space with confinement with
chiral symmetry breaking}

As we said above, for $N_f \ge N_c$ the vacuum degeneracy can not be
lifted.  Therefore, the quantum theory also has a continuous space of
inequivalent vacua.  Since this space can be different than the
classical one, we will refer to it as the ``quantum moduli space.''
The most interesting questions about it are associated with the nature
of its singularities.  Classically, the singularities were associated
with massless gluons.  Are there singularities in the quantum moduli
space?  What massless particles are associated with those
singularities?

The classical moduli space for $N_f=N_c$ is given in terms of the gauge
invariant description \flatmbtb\ as the the space of expectation
values of mesons $M^{i}_{\tilde j}$,
and baryons $B$ and $\widetilde B$ subject to the
classical constraint
\eqn\nfnccl{ \det M - \widetilde B B =0,}
which follows from Bose statistics of $Q$ and $\widetilde Q$. This
space has a singular submanifold $B=\widetilde B=0$ and rank$(M)\leq
N_c-2$, where $d(\det M-B\widetilde B)=0$.  Physically, the classical
singularities reflect the fact that there are additional degrees of
freedom, $SU(N_c-\hbox{rank}(M))$ gluons, on this submanifold.

The quantum moduli space is parameterized by the same fields but the
constraint is modified \nati\ to
\eqn\sqmsc{ \det M - \widetilde B B =\Lambda^{2N_c}.}
This can be seen by adding mass terms $W_{\rm tree}=\Tr mM$ and taking
$m\rightarrow 0$.  It follows from \mevm\ that $\det \ev{M}=\Lambda
^{2N_c}$, independent of $m$, for $N_f=N_c$.  This agrees with \sqmsc\
for $\ev{B\widetilde B}=0$, which is the case with only added mass
terms.  Considering more generally $W_{\rm tree}=\Tr m M+bB+\tilde
b\widetilde B$ with $m,b,\tilde b\rightarrow 0$, $\ev{B\widetilde B}$
can be non-zero and the expectation values are found to satisfy
\sqmsc\intin.
Because the right hand side of \sqmsc\ is proportional to the
one-instanton action, the quantum modification of the classical
constraint is exactly given by a one-instanton contribution.

There are no singularities on the quantum moduli space given by
\sqmsc\ -- all of the classical singularities have been smoothed out
by quantum effects.  As a similar but simpler example of the deformed
moduli space, consider the space $XY=\mu$ in $C^2$.  For $\mu=0$ the
space is a pair of cones (corresponding to the $X$ plane and the $Y$
plane) touching at their tips; the space is singular at the origin,
where the cones touch.  In the deformed space $\mu \neq 0$, the two
cones are smoothed out to a single sheeted hyperboloid.  It
asymptotes to the two cones far from the origin but
has a smooth hourglass shape where they connect.  Similarly, the
quantum space \sqmsc\ is smooth, which is
dramatically different from the classical
space \nfnccl\ near the origin.  For large expectation values
$M$, $B$ and $\widetilde B$, the difference becomes negligible, as it
should be in the weak coupling region.

Since there are no singularities on the quantum moduli space, the only
massless particles are the moduli, the fluctuations of $M$, $B$, and
$\widetilde B$ preserving \sqmsc.  In the semi-classical region of large
fields it is appropriate to think of the theory as ``Higgsed.''  Near
the origin, because the quantum theory is smooth in terms of the
mesons and baryons, it is appropriate to think of the theory as being
confining.  There is a smooth transition from the region where a Higgs
description is more appropriate to the region where a confining
description is more appropriate.  Again, this is possible because of
the presence of matter fields in the fundamental representation of the
gauge group \higgscon.

Because the origin $M=B=\widetilde B=0$ is not on the quantum moduli
space
\sqmsc, the
quantum dynamics necessarily break the anomaly free, chiral
$SU(N_f)\times SU(N_f)\times U(1)_B\times U(1)_R$ symmetry in
\globsymc.  Different points on the quantum moduli space exhibit
different patterns of chiral symmetry breaking.  For example at
$M^{i}_{\tilde j} =\Lambda^2
\delta^{i}_{\tilde j},$ $B=\widetilde B=0$ the symmetry is broken as
\eqn\xsbi{SU(N_f)_L\times
SU(N_f)_R \times U(1)_B \times U(1)_R \rightarrow
SU(N_f)_V \times U(1)_B \times U(1)_R .}
At $M=0,$ $B=- \widetilde B = \Lambda^{N_c}$ the breaking pattern is
\eqn\xsbii{SU(N_f)_L\times
SU(N_f)_R \times U(1)_B \times U(1)_R \rightarrow
SU(N_f)_L\times SU(N_f)_R \times U(1)_R .}  Some of the moduli are
Goldstone bosons of the broken symmetry whereas others take the theory
to vacua with different breaking patterns.  For no vacuum in the
quantum moduli space \sqmsc\ is the full chiral symmetry unbroken.  It
is straightforward to check that the massless fermion spectrum,
consisting of the fermionic components of the chiral superfield
moduli, satisfies the 'tHooft anomaly conditions for the unbroken
symmetries.

The constraint \sqmsc\ can be implemented with a superpotential
$W=A(\det M-B\widetilde B-\Lambda ^{2N_c})$, with $A$ a Lagrange
multiplier.  The reader can verify that, upon adding $W_{\rm
tree}=mM^{N_c}_{\tilde N_c}$ to give a mass to the $N_c$-th flavor,
the low energy theory with $N_f=N_c-1$ light flavors has the
appropriate superpotential \wsymi.

\subsec{$N_f = N_c+1$ -- Confinement without chiral
symmetry breaking}

We now add another
massless flavor to the previous case.  The classical moduli
space is again described as in \flatmbtb\
by the mesons $M$, baryons $B_i=
\epsilon _{ij_1\dots j_{N_c}}Q^{j_1}\dots Q^{j_{N_c}}$ and
$\widetilde B^{\tilde i}=\epsilon ^{\tilde i\tilde j_1\dots \tilde
j_{N_c}}\widetilde Q_{\tilde j_1}\dots \widetilde Q_{\tilde j_{N_c}}$
subject to the constraints
\eqn\nfncpic{ \eqalign{&
\det M \left( {1 \over M} \right)_{i}
^{\tilde j} - B_i\widetilde B^{\tilde j} =0 \cr &M^{i}_{\tilde j} B_i=
M^{i}_{\tilde j}\widetilde B^{\tilde j}=0 .\cr} }

Unlike the previous case, for $N_f=N_c+1$ the quantum moduli space is
the same as the classical one \nati.  This can be seen by adding
$W_{\rm tree}=\Tr mM$, giving
masses to all $N_f+1$ flavors.  The
expectation values of $M$ are then given by \mevm.  By taking the limit
$m_{i}^{\tilde j}\rightarrow 0$ in different ratios, it is found that
$\ev{M^{i}_{\tilde j}}$ can be anywhere on the classical moduli
space \nfncpic\ of
vacua.

Because the quantum moduli space of vacua is the same as the classical
moduli space, it has singularities at strong coupling.  The
singularities, however, are interpreted differently than in the
classical theory.  Rather than being associated with massless gluons,
the singularities are associated with additional massless mesons and
baryons!  In particular, at the point $M=\widetilde B = B=0$ the
global chiral symmetry \eqn\unbso{ SU(N_f)_L\times SU(N_f)_R \times
U(1)_B
\times U(1)_R} is unbroken and all the components of $M$, $B$ and
$\widetilde B$ are massless and physical \nati.  It is a non-trivial
consistency check that this massless spectrum at the origin satisfies
the 'tHooft anomaly matching conditions for the full global symmetry
\unbso.

Away from the origin, all the degrees of freedom in $M$, $B$ and
$\widetilde B$ are physical and they couple through the superpotential
\nati
\eqn\cwocsb{W_{\rm eff}= {1
\over \Lambda^{2N_c -1 }}(M^{i}_{\tilde j}B_i
\widetilde B^{\tilde j} - \det M ). }
The classical constraints \nfncpic\ appear as the equations of motion
${\partial W_{\rm eff} \over \partial M}= {\partial W_{\rm eff} \over
\partial B}= {\partial W_{\rm eff} \over \partial \widetilde B}=0$.
Far {}from the origin, in the region of weak coupling, the number of
independent massless fields is the same as in the classical theory
because the components of $M$, $B$, and $\widetilde B$ which are
classically constrained get a large mass from \cwocsb.

We conclude that the spectrum at the origin of field space consists of
massless composite mesons and baryons and that the chiral symmetry of
the theory is unbroken there.  This is confinement without chiral
symmetry breaking.  Again, we see a smooth transition \higgscon\ from
the semi-classical region where a Higgs description is more appropriate
to a strongly coupled region where a confining description is more
appropriate.

The reader can verify that adding $W_{\rm tree}=mM^{N_c+1}_{
\widetilde{N_c+1}}$ to \cwocsb\
to give a mass to the $N_c+1$-th flavor yields the quantum moduli
space with constraint \sqmsc\ in the low energy theory with $N_f=N_c$
light flavors.

\newsec{SUSY QCD for $N_f>N_c+1$}

\subsec{$N_f\geq 3N_c$}

In this range the theory is not asymptotically free.  This means that,
because of screening, the coupling constant becomes smaller at large
distances.  Therefore, the spectrum of the theory at large distance can
be read off {}from the Lagrangian -- it consists of the elementary
quarks and gluons.  The long distance behavior of the potential between
external electric test charges has the free electric behavior in \wlp.
For this range of $N_f$, the theory is in a non-Abelian free electric
phase.

We should add here that, strictly speaking, such a theory is not well
defined as an interacting quantum field theory because of the
Landau pole at $R\sim \Lambda ^{-1}$. However, it can be a
consistent description of the low energy limit of another theory.

\subsec{${3 \over 2}N_c < N_f < 3N_c$;
Interacting non-Abelian Coulomb phase}

In this range the theory is asymptotically free.  This means that at
short distance the coupling constant is small and it becomes larger at
longer distances.  However, for this range of $N_f$
\refs{\nati, \sem}, rather
than growing to infinity, it reaches a finite value -- a fixed point of
the renormalization group.

The exact
beta function in supersymmetric QCD satisfies \refs{\rusano,\sv}
\eqn\betafun{\eqalign{
\beta(g)&= - {g^3 \over 16 \pi^2} {3N_c - N_f + N_f \gamma(g^2) \over 1
- N_c {g^2 \over 8 \pi^2}} \cr
\gamma(g^2) & = -{g^2 \over 8 \pi^2}{N_c^2 -1 \over N_c} + \CO( g^4),
\cr}}
where $\gamma(g^2)$ is the anomalous dimension of the mass.  Since
there are values of $N_f$ and $N_c$ where the one loop beta function
is negative but the two loop contribution is positive, there might be
a non-trivial fixed point
\ref\bankszaks{T. Banks and A. Zaks, \np{196}{1982}{189}}.
Indeed, by taking $N_c$ and $N_f$ to infinity holding $N_c g^2$ and
${N_f \over N_c}=3 - \epsilon$ fixed, one can establish the existence
of a zero of the beta function at $N_cg_*^2={8 \pi^2 \over 3} \epsilon
+
\CO(\epsilon^2) $.  Therefore, at least for large $N_c$ and $\epsilon=
3-{N_f \over N_c} \ll 1$, there is a non-trivial fixed point.  It was
argued in \sem\
that such a fixed point exists for every ${3\over 2}N_c < N_f < 3N_c$.

Therefore, for this range of $N_f$, the infrared theory is a
non-trivial four dimensional superconformal field theory.  The
elementary quarks and gluons are not confined but appear as
interacting massless particles.  The potential between external
electric sources behaves as $$V \sim {1 \over R}$$ and therefore we
refer to this phase of the theory as the non-Abelian Coulomb phase.

Given that such a fixed point exists, we can use the superconformal
algebra to derive some exact results about the theory.  This algebra
includes an $R$ symmetry.  It follows {}from the algebra that the
dimensions of the operators satisfy
\eqn\dimchir{D \ge {3 \over 2} |R|;}
the inequality is saturated for chiral operators, for which $D = {3
\over 2} R $, and for anti-chiral
operators, for which $D =- {3 \over 2}
R $.  Exactly as in N=2 theories in two dimensions, this has
important consequences.  Consider the operator product of two chiral
operators, $\CO_1(x) \CO_2(0)$.  All the operators in the resulting
expansion have $R=R(\CO_1) +R(\CO_2)$ and hence $D\ge
D(\CO_1)+D(\CO_2)$.  Therefore, there is no singularity in the
expansion at $x=0$ and we can define the product of the two operators
by simply taking the limit of $x$ to zero.  If this limit does not
vanish, it leads to a new chiral operator $\CO_3$ whose dimension is
$D(\CO_3)=D(\CO_1)+D(\CO_2)$.  We conclude that the chiral operators
form a ring.

The $R$ symmetry of the superconformal fixed point is not anomalous
and commutes with the flavor $SU(N_f) \times SU(N_f)\times U(1)_B$
symmetry.  Therefore, it must be the anomaly free $R$ symmetry
appearing in \qtransl.  Hence the gauge invariant operators
$\widetilde Q Q$ have \sem\
\eqn\rdqtq{D(\widetilde Q Q)
= {3 \over 2} R(\widetilde Q Q) = 3 {N_f-N_c \over
N_f}} and similarly
\eqn\brdqtq{D(B)=D (\widetilde B)= {3N_c(N_f-N_c) \over 2N_f}.}
The value of $D(\widetilde Q Q)$ also follows {}from \betafun\
-- at the zero of the beta function $\gamma=-3{N_c \over N_f} +1$ and
hence $D=\gamma+2= 3 {N_f-N_c \over N_f}$.

All of the gauge invariant operators at the infrared
fixed point should be
in unitary representations of the superconformal algebra.
The complete list of such representations was given in
\ref\supconfra{M. Flato and C. Fronsdal, Lett. Math. Phys. {\bf 8}
(1984) 159; V.K. Dobrev and V.B. Petkova, \pl{162}{1985}{127}}
by extending the analysis
\ref\mack{G. Mack, \cmp{55}{1977}{1}} of
the ordinary conformal algebra.
One of the constraints on the representations
which follows already {}from the analysis of \mack\ is that spinless
operators have $D \ge 1$ (except the identity operator with $D=0$) and
the bound is saturated for free fields (satisfying $\partial_\mu
\partial^\mu \Phi =0$).  For $D<1$ ($D\not= 0$) a highest weight
representation includes a negative norm state which cannot exist in a
unitary theory.

The fixed point coupling $g_*$ gets larger as the number of flavors is
reduced.  For $N_f$ at or below ${3\over 2}N_c$ the theory is very
strongly coupled and goes over to a new phase, different from the
interacting non-Abelian Coulomb phase.  To see that the theory must be
in a different phase, note that the value of $D(\widetilde Q Q)$ in
\rdqtq\ is inconsistent with the unitarity bound $D\geq 1$ for $N_f <
{3\over 2}N_c$. The new phase will be explained below.  A clue is the
fact that the dimension of $M=\widetilde Q Q$ becomes one for
$N_f={3\over 2}N_c$, which shows that $M$ becomes a free field, i.e.
$\partial^2 M=0$.  This suggests that in the correct description for
$N_f={3\over 2}N_c$ the field $M$, and perhaps even the whole IR
theory, is free.

\subsec{Duality}

The physics of the interacting fixed point obtained
for the range ${3\over 2}N_c<N_f<3N_c$ has an equivalent,
``magnetic,'' description
\sem.  It is based on the gauge group
$SU(N_f-N_c)$, with $N_f$ flavors of quarks $q_i$ and $\widetilde
q^{\tilde i}$ and gauge invariant fields $M^{i}_{\tilde j}$ with a
superpotential
\eqn\wmagi{W={1\over \mu}M^{i}_{\tilde j}q_i\widetilde q^{\tilde j}.}
We will refer to this gauge group as the magnetic gauge group
and to its quarks as magnetic quarks.  Without the superpotential
\wmagi, the magnetic theory  also flows to a
non-Abelian Coulomb phase fixed point because ${3 \over 2}(N_f-N_c) <
N_f < 3(N_f-N_c)$ for the above range of $N_f$.  At this fixed point
$M$ is a free field of dimension one and, using \rdqtq ,
$D(q\widetilde q)=3N_c/N_f$.  Because the dimensions of chiral
operators add, the superpotential \wmagi\ has dimension
$D=1+3N_c/N_f<3$ at the fixed point of the magnetic gauge theory and
is thus a relevant perturbation, driving the theory to a new fixed
point.  The surprising fact is that this new fixed point is identical
to that of the original, ``electric,'' $SU(N_c)$ theory.  Note that
the two theories have different gauge groups and different numbers of
interacting particles.  Nevertheless, they describe the same fixed
point.  In other words, there is no experimental way to determine
whether the $1
\over R$ potential between external sources is mediated by the
interacting electric or the interacting magnetic variables.  Such a
phenomenon of two different Lagrangians describing the same long
distance physics is common in two dimensions and is known there as
quantum equivalence.  These four dimensional examples generalize the
duality \mo\ in finite $N=4$ supersymmetric theories \dualnf\ and in
finite $N=2$ theories \swii\ to asymptotically free $N=1$ theories.

The scale $\mu$ in \wmagi\ is needed for the following reason.  In the
electric description $M^{i}_{\tilde j}=Q^i \widetilde Q_{\tilde j}$ has
dimension two at the UV fixed point and acquires anomalous dimension
\rdqtq\ at the IR fixed point.  In the magnetic description, $M_m$ is
an elementary field of dimension one at the UV fixed point which flows
to the same operator with dimension \rdqtq\ at the IR fixed point.  In
order to relate $M_m$ to $M$ of the electric description in the UV, a
scale $\mu$ must be introduced with the relation $M=\mu M_m$.  Below
we will write all the expressions in terms of $M$ and $\mu$ rather
than in terms of $M_m$.

The magnetic theory has a scale $\widetilde
\Lambda$ which is related to
the scale $\Lambda$ of the electric theory by
\eqn\mgscrgen{\Lambda^{3N_c-N_f}\widetilde
\Lambda^{3(N_f-N_c)-N_f}=(-1)^{N_f-N_c}\mu ^{N_f},}
where $\mu$ is the dimensionful scale explained above.  This relation
of the scales has several consequences:

\item{1.}
It is easy to check that it is preserved under mass deformations and
along the flat directions (more details will be given below).  The
phase $(-1)^{N_f-N_c}$ is important in order to ensure that this is
the case.

\item{2.} It shows that
as the electric theory becomes stronger the magnetic
theory becomes weaker and vice versa.  It is the analog of
$g\rightarrow 1/g$ for asymptotically free theories.

\item{3.}  Differentiating the action with respect to $\log \Lambda$
relates the field strengths of the electric and the magnetic theories
as $W_\alpha^2 = -\widetilde W_\alpha^2$.  The minus sign in this
expression is common in electric magnetic duality, which maps
$E^2-B^2=-(\widetilde E^2-\widetilde B^2)$.  In our case it shows that
the gluino bilinear in the electric and the magnetic theories are
related by $\lambda\lambda=-\tilde \lambda\tilde \lambda$.

\item{4.}
Because of the phase $(-1)^{N_f-N_c}$, the relation \mgscrgen\ does
not look dual -- if we perform another duality transformation it
becomes $\Lambda^{3N_c-N_f}\widetilde
\Lambda^{3(N_f-N_c)-N_f}=(-1)^{N_c}
\tilde \mu ^{N_f}$ and therefore
\eqn\mumutil{\tilde \mu=-\mu .}
This minus sign is important when we dualize again.  The dual of the
dual magnetic theory is an $SU(N_c)$ theory with scale $\Lambda$,
quarks $d^i$ and $\widetilde d_{\tilde j}$, and additional singlets
$M^{i}_{\tilde j}$ and $N_{i}^{\tilde j}=q_i \widetilde q^{\tilde j}$,
with superpotential
\eqn\typsupa{W={1 \over \tilde \mu} N_{i}^{\tilde
j}d^i\widetilde d_{\tilde j}+{1 \over \mu} M^{i}_{\tilde
j}N_{i}^{\tilde j}= {1 \over \mu} N_{i}^{\tilde j}(-d^i \widetilde
d_{\tilde j}+M^{i}_{\tilde j}) .}  The first term is our standard
superpotential of duality transformations\foot{The relative minus sign
between it and
\wmagi, which follows from \mumutil, is common in Fourier or Legendre
transforms. (Compare with \invlt\ and \wl.)} and
the second term is simply copied from \wmagi.  $M$ and
$N$ are massive and can be integrated out using their equations of
motion: $N=0$, $M^{i}_{\tilde j}=d^i\widetilde d_{\tilde j}$.
This last relation shows that
the quarks $d$ and $\widetilde d$
can be identified with the original electric quarks
$Q$ and $\widetilde Q$.  The dual of the magnetic theory is
the original electric theory.

The electric and magnetic theories have different gauge symmetries.
This is possible because gauge symmetries really have to do with a
redundant description of the physics rather than with symmetry.  There
is no problem with having two different redundant descriptions of the
same physics.  On the other hand, global symmetries are physical and
should be the same in the electric and magnetic theories.  Indeed, the
magnetic theory has the same anomaly free global $SU(N_f)\times
SU(N_f)\times U(1)_B\times U(1)_R$ as the electric theory, with the
singlet $M^{i}_{\tilde j}$ transforming as $Q^i\widetilde Q_{\tilde
j}$ and the magnetic quarks transforming as
\eqn\newquaqu{\eqalign{
q \qquad {\rm in} \qquad &(\overline N_f,1,{N_c \over N_f-N_c},
{N_c\over N_f}) \cr
\widetilde q \qquad {\rm in} \qquad &(1, N_f,- {N_c \over N_f-N_c},
{N_c\over N_f}) . \cr }}
This symmetry is anomaly free in the magnetic theory and it is
preserved by the
superpotential \wmagi.
Furthermore, the magnetic spectrum with these charges
satisfies the 'tHooft anomaly matching conditions.

In order for the dual theory to describe the same physics as the
electric theory, there must be a mapping of all gauge invariant
primary operators of the electric theory to those of the dual theory.
For example, the electric mesons $M^{i}_{\tilde j}=Q^i\widetilde
Q_{\tilde j}$ and the singlets $M^{i}_{\tilde j}$ of the magnetic
theory become identical in the infrared.  All such mappings must be
compatible with the global symmetry charges discussed above.  Another
set of gauge invariant operators of the electric theory are the
baryons $B^{i_1\dots i_{N_c}}=Q^{i_1}\cdots Q^{i_{N_c}}$ and
$\widetilde B_{\tilde i_1\dots
\tilde i_{N_c}}=\widetilde Q_{\tilde i_1}\cdots
\widetilde Q_{\tilde i_{N_c}}$.
In the magnetic theory we can similarly form the baryons $b_{i_1\dots
i_{\tilde N_c}}=q_{i_1}\cdots q_{i_{\tilde N_c}}$ and $\widetilde
b^{\tilde i_1\dots \tilde i_{\tilde N_c}}=\widetilde q^ {\tilde
i_1}\cdots \widetilde q^{\tilde i_{\tilde N_c}}$, where $\widetilde
N_c=N_f-N_c$. At the fixed point, these operators are related via
\eqn\barmap{\eqalign{B^{i_1\dots i_{N_c}}&=C
\epsilon ^{i_1\dots
i_{N_c}j_1\dots j_{\tilde N_c}}
b_{j_1\dots j_{\tilde N_c}},\cr \widetilde
B_{\tilde i_1\dots \tilde i_{N_c}}&=
C\epsilon _{\tilde i_1\dots
\tilde i_{N_c}\tilde j_1\dots \tilde j_{\tilde N_c}}\widetilde
b^{\tilde j_1\dots \tilde j_{\tilde N_c}},}} with
$C=\sqrt{-(-\mu)^{N_c-N_f}\Lambda ^{3N_c-N_f}}$.  Note that these
mappings respect the global symmetries discussed above.  The
normalization constant $C$ was fixed by symmetries and by limits to be
discussed below.  It follows from \mgscrgen\ that \barmap\
respects the $Z_2$ nature of the duality.

\subsec{$N_c+2\leq N_f\leq {3\over 2}N_c$}

Recall that the electric $SU(N_c)$ theory with $N_f$ flavors was in a
non-Abelian free electric phase for $N_f\geq 3N_c$ and in the
non-Abelian Coulomb phase for ${3\over 2}N_c<N_f<3N_c$, with the fixed
point at larger electric coupling for smaller $N_f$.  In the magnetic
description of the non-Abelian Coulomb phase fixed point, the magnetic
coupling at the fixed point is small for $N_f$ near $3(N_f-N_c)$ and
gets larger with $3(N_f-N_c)-N_f$; in the magnetic description the
fixed point is at weaker coupling for smaller $N_f$.  It was seen that
for $N_f\leq {3\over 2}N_c$ the theory must be in a different phase.
In the magnetic description, the situation is clear: since $3(N_f-N_c)
\le N_f$, the magnetic $SU(N_f-N_c)$ gauge theory is not
asymptotically free (and the added superpotential \wmagi\ is
irrelevant) and thus weakly coupled at large distances.  Therefore,
the low energy spectrum of the theory consists of the $SU(N_f-N_c)$
gauge fields and the fields $M$, $q$, and $\widetilde q$ in the dual
magnetic Lagrangian
\sem.  These magnetic massless states are composites of the elementary
electric degrees of freedom.  The massless composite gauge fields
exhibit gauge invariance which is not visible in the underlying
electric description.  The theory generates new gauge invariance!
Because there are massless magnetically charged fields, the theory is
in a non-Abelian free magnetic phase.

\subsec{Deformations of the dual theories}
We now consider deforming the theories, showing in detail how the
electric-magnetic duality exchanges strong coupling with weak
coupling and Higgs with confinement.

Consider starting from the electric theory and giving a mass to the
$N_f$-th flavor by adding $W_{\rm tree}=mM^{N_f}_{\tilde N_f}$.  The
low energy theory has $N_f-1$ light flavors and a scale $\Lambda _L$
related to the scale $\Lambda$ of the high energy theory by $\Lambda
_L^{3N_c-(N_f-1)}=m\Lambda ^{3N_c-N_f}$.  As expected, the low energy
electric theory is at stronger coupling; the mass term is a relevant
operator, sending the theory to a more strongly coupled fixed point.

In the magnetic theory, adding $W_{\rm tree}$ gives
\eqn\massdefd{W={1\over \mu}M^i_{\tilde i} q_i
\widetilde q^{\tilde i} +
m M^{N_f}_{\tilde N_f} .}  The equations of motion of $
M^{N_f}_{\tilde N_f}$, $ M^{i}_{\tilde N_f}$ and $ M^{N_f}_{\tilde i}$
lead to
\eqn\massdeme{
q_{N_f} \widetilde q^{\tilde N_f} = -\mu m,\qquad q_{i} \widetilde
q^{\tilde N_f} = q_{N_f} \widetilde q^{\tilde i} = 0 ,} (color indices
are suppressed and summed) which show that the magnetic gauge group is
broken by the Higgs mechanism to $SU(N_f-N_c-1)$ with $N_f-1$ quarks
left massless.  The equations of motion of the massive quarks lead to
\eqn\massdeqe{M^{N_f}_{\tilde N_f}=M^{i}_{\tilde N_f}=
M^{N_f}_{\tilde i}=0 .}
The low energy superpotential is
\eqn\lowensuph{W={1\over \mu}\widehat
M^i_{\tilde i} \hat q_i \hat {\widetilde q}{}^{\tilde i},
\qquad i,\tilde i = 1,...,N_f-1,}
where $\widehat M$, $\hat q$ and $\hat{ \widetilde q}$ are the light
fields with $N_f-N_c-1$ colors and $N_f-1$ flavors.  The scale of the
low energy magnetic theory is given by $\widetilde \Lambda
_L^{3(\widetilde N_c-1)-(N_f-1)}=\widetilde
\Lambda ^{3\widetilde N_c-N_f}/\ev{q_{N_f}
\widetilde q^{\tilde N_f}}$.  Note
that \mgscrgen\ and \barmap\ are preserved in the low energy theories.
The low energy magnetic theory is at weaker coupling and is the dual
of the low energy electric theory.  The duality is preserved under the
mass term deformation and exchanges a more strongly coupled electric
description of the new fixed point with a more weakly coupled magnetic
description of the new fixed point.

The above discussion of the mass term deformation is incomplete for
$N_f=N_c+2$, where the mass term for the $(N_c+2)$-th flavor triggers
complete breaking of the magnetic gauge group.  The low energy theory
contains the mesons $\widehat M_{\tilde i}^j$, where the hat refers to
the flavors $i, \tilde i=1\dots N_c+1$, and the singlets $\hat q_i$
and $\hat{\widetilde q}{} ^{\tilde i}$, which are the components of
the magnetic quarks remaining massless after the Higgs mechanism, with
a superpotential as in \lowensuph.  The map \barmap\ for $N_f=N_c+1$
implies that the singlets $\hat q_i$ and $\hat {\widetilde q}{}
^{\tilde i}$ are, in fact, the baryons $B_i$ and $\widetilde B^{\tilde
i}$ of the low energy electric theory.  It has already been suggested,
at least for large $N_c$, that baryons can be thought of as solitons
in the pion Lagrangian
\ref\barsol{T.H.R. Skyrme, Proc.Roy.Soc. {\bf A260} (1961) 127; E.
Witten, \np {160}{1979}{57}; \np{223}{1983}{422}; \np
{223}{1983}{433}}.  Here we see an explicit realization of a related
idea -- the baryons are magnetic monopoles of the elementary quarks
and gluons!  Taking the normalization in \barmap\ into account, the
superpotential \lowensuph\
obtained in the low energy theory from flowing down
{}from that of the magnetic, $N_f=N_c+2$, $SU(2)$ theory is
\eqn\woi{W_o={1\over \Lambda _L^{2N_c-1}}\widehat
M^i_{\tilde j}B_i\widetilde B^{\tilde j},} where $\Lambda _L$ is the
scale of the low energy electric theory with $N_f=N_c+1$ light
flavors.  However, because the magnetic theory is completely Higgsed
in the flow down from $N_f=N_c+2$, the low energy theory also includes
instanton contributions in the broken magnetic group.  In particular,
the superpotential has an additional term which is the magnetic analog
of \wsymi\
\eqn\maginst{W_{inst}={\widetilde \Lambda _L^{6-(N_c+2)}\det
(\mu ^{-1}\widehat
M)\over q^{N_f+2}\widetilde q^{N_f+2}}=-{\det \widehat M\over \Lambda
_L^{3N_c-(N_c+1)}},}
where  use was made of
\mgscrgen.  Adding this to the superpotential
\woi\ and dropping the hats and the $L$ subscript, the
superpotential of the low energy theory with $N_f=N_c+1$ light flavors
is
\eqn\wtrmaginst{W={1\over \Lambda ^{2N_c-1}}(
M^i_{\tilde j}B_i\widetilde B^{\tilde j} -\det M).}  This is precisely
the superpotential
\cwocsb.  In the electric description \cwocsb\ is
associated with strong coupling effects.  In the magnetic description
it is thus rederived in a weakly coupled framework.

Another way to analyze the theory with mass terms is to consider the
massless theory for generic values of $\ev{M}$.  The dual quarks
acquire mass $\mu ^{-1} M$ and the low energy magnetic theory is pure
glue $SU(N_f-N_c)$ Yang-Mills theory with scale $\widetilde
\Lambda_L^{3(N_f-N_c)}=
\mu^{-N_f}\widetilde\Lambda^{3(N_f-N_c)-N_f} \det
M$.  As in \ymw, gluino condensation in this theory leads to an
effective superpotential
\eqn\effsupgen{
W_{\rm eff}=(N_f-N_c)\widetilde \Lambda_L^3=
(N_c-N_f)\left({\Lambda^{3N_c-N_f}\over
\det M}\right)^{1/(N_c-N_f)},}
where we used \mgscrgen\ (the $(-1)^{N_f-N_c}$ sign in that relation
plays a crucial role in getting the correct overall sign here), which
is the same as the continuation of
\wsymi\ to these values of $N_c$, $N_f$.  This guarantees that the
superpotentials \wsymi\ and the expectation values of $\ev{M^i_{\tilde
j}}$ are reproduced correctly when mass terms are added to the
magnetic theory.

The infrared fixed point can also be deformed by
giving the fields in the electric theory expectation values
along the $D$-flat directions \flatqii.
Consider, for example, large $\ev{Q^{N_f}}=\ev{\widetilde Q_{
\tilde N_f}}$, breaking the
electric $SU(N_c)$ theory with $N_f$ flavors to $SU(N_c-1)$ with
$N_f-1$ light flavors. The low energy electric theory has scale
$\Lambda _L^{3(N_c-1)-(N_f-1)}=\Lambda
^{3N_c-N_f}/\ev{Q^{N_f}\widetilde Q_{N_f}}$ and the fixed point is at
weaker coupling.  In the magnetic description, the large
$\ev{M^{N_f}_{\tilde N_f}}$ gives a large mass $\mu
^{-1}\ev{M^{N_f}_{\tilde N_f}}$ to the flavor $q_{N_f}\widetilde
q^{\tilde N_f}$.  The low energy magnetic theory is $SU(N_f-N_c)$ with
$N_f-1$ light flavors and scale $\widetilde \Lambda
_L^{3(N_f-N_c)-(N_f-1)}=\mu ^{-1}\ev{M^{N_f}_{\tilde N_f}}\widetilde
\Lambda ^{3(N_c-N_f)-N_f}$.  The low energy magnetic theory is at
stronger coupling and is the dual of the low energy electric theory.
Note that the flow preserves \mgscrgen\ and \barmap.  Deformations
along the flat
directions with $\ev{B}\neq 0$ directions were analyzed in
\ref\aharony{O. Aharony, hep-th/9502013, \pl{351}{1995}{220}.}.

Classically, and to all orders of perturbation theory, the electric
and magnetic theories have different moduli spaces of vacua -- it is
only after taking non-perturbative effects into account that they are
seen to be identical.  For example, in the electric theory there is a
classical constraint rank$\ev{M}\leq N_c$.  In the dual theory, $M$ is
an independent field whose expectation value is unconstrained to all
orders of perturbation theory -- the constraint arises in the dual
theory by quantum effects!  Defining $N_i^{\tilde j}=q_i\widetilde
q^{\tilde j}$, the $M$ equations of motion imply that vacua of \wmagi\
are at $\ev{N_i^{\tilde j}}=0$.  However, because the magnetic theory
has $N_f-{\rm rank}\ev{M}$ massless flavors, for $N_f-{\rm
rank}\ev{M}<N_f-N_c$ the magnetic theory generates a superpotential
analogous to \wsymi\ and there is no vacuum with $\ev{N}=0$.  The
vacua of the dual theory thus also satisfy rank$\ev{M}\leq N_c$ but as
a result of quantum effects rather than as a classical constraint.
Similarly, for ${\rm rank}\ev{M}=N_c$, the magnetic theory has
$N_f-N_c=\widetilde N_c$ massless fields and thus develops a
constraint analogous to \sqmsc
\eqn\sqmscmd{\det N-b\widetilde b=\widetilde \Lambda _L^{2
\widetilde N_c},}
where $\widetilde \Lambda _L^{2\widetilde N_c}=
\det ' \ev{\mu ^{-1}M}\widetilde
\Lambda ^{3\widetilde N_c-N_f}$, with $\det '\ev{\mu ^{-1}
M}$ the product of the $N_c$ non-zero eigenvalues of $\ev{\mu
^{-1}M}$.  Using the $M$ equation of motion, $\ev{N}=0$, and the
mapping \barmap\ and \mgscrgen, the relation \sqmscmd\ yields
\eqn\clfquii{\ev{B\widetilde B}=\det{}' \ev{M}.}
This derives a constraint which is classical in the electric
description (it was discussed following \flatqii ) via
quantum dynamics in the dual theory.   In
addition to providing a necessary check on the duality, the fact that
classical relations arise via quantum effects in the dual is
interesting in that, conversely, quantum effects in one theory can be
related to classical identities in the dual.

\newsec{$SO(N_c)$ with $N_f \times {\bf N_c}$}

In the $SU(N_c)$ theories discussed above, with matter in the
fundamental representation, there is no invariant distinction between
Higgs and confinement \higgscon.  This is not the case in theories
based on $SO(N_c)$ with $N_f
\times {\bf N_c}$ because the matter cannot screen sources
in the spinor representation \foot{We will
limit our considerations to the
Lie algebra and not distinguish between $SO(N_c)$ and
Spin$(N_c)$}.  These theories, therefore, lead to a
clearer picture of the dynamics.  In particular, here the transition
{}from the Higgs phase to the Confining phase occurs with a well
defined phase transition.

Many of the results in these $SO(N_c)$ theories
\refs{\ads, \ils, \swi, \intse, \sem, \isson}
are similar to the results in $SU(N_c)$, showing that some phenomena
are generic.  These $SO(N_c)$ theories also exhibit many new
phenomena, which are not present in the $SU(N_c)$ examples.  The most
dramatic of them is oblique confinement
\nref\thooft{G. 'tHooft, \np{190}{1981}{455}.}%
\nref\cardyrabin{J. Cardy and E. Rabinovici, \np{205}{1982}{1};
J. Cardy, \np{205}{1982}{17}.}%
\refs{\thooft,\cardyrabin}, driven by the condensation of dyons
(particles with both electric and magnetic charges).
This phenomenon is best described by another equivalent theory -- a
dyonic theory.  Therefore, these theories exhibit
electric-magnetic-dyonic triality \refs{\isson, \isprev}.
The discussion here will be brief; many more details can be found in
\isson.

\subsec{The phases}

For $N_f\geq 3(N_c-2)$ massless quarks $Q$ in the $N_c$ dimensional
representation of $SO(N_c)$, the theory is not asymptotically free and
the infrared theory is thus in a free electric phase.  For ${3\over
2}(N_c-2)<N_f<3(N_c-2)$, the theory is asymptotically free and flows to
a non-Abelian Coulomb phase fixed point in the infrared.  This phase
has a dual magnetic description in terms of an $SO(N_f-N_c+4)$ gauge
theory which we review below.  For $N_c-2<N_f\leq {3\over 2}(N_c-2)$,
the theory is in a free magnetic phase with a dynamically generated
composite $SO(N_f-N_c+4)$ gauge invariance.  For $N_f=N_c-2$, the
theory is in an Abelian Coulomb phase.

\subsec{Duality }

As discussed in \refs{\sem, \isson}, the infra-red behavior of these
theories has a dual, magnetic description in terms of an
$SO(N_f-N_c+4)$ gauge theory with $N_f$ flavors of dual quarks $q_i$
and an additional gauge singlet field $M^{ij}=Q^i\cdot Q^j$ with a
superpotential
\eqn\typsup{W={1\over 2\mu}M^{ij}q_i\cdot q_j}
(an additional term is required for $N_f=N_c-1$).  The scale $\mu$ is
needed for the same reason as in the $SU(N_c)$ duality.
For generic $N_c$ and $N_f$ the scale $\widetilde \Lambda$
of the magnetic theory is related to the scale
$\Lambda$ of the electric theory by
\eqn\mgscrgenn{2^8\Lambda^{3(N_c-2)-N_f}\widetilde
\Lambda^{3(N_f-N_c+2)-N_f}=(-1)^{N_f-N_c}\mu ^{N_f}.}
The constant $2^8$ reflects the $\overline{DR}$ conventions\foot{
In these
conventions the matching relation between the scale of the high energy
theory with $N_f$ flavors and the mass term $W_{tree}=\half
mQ^{N_f}\cdot Q^{N_f}$ and the low energy theory with $N_f-1$ flavors
is $\Lambda _L^{3(N_c-2)-N_f+1}=m\Lambda ^{3(N_c-2)-N_f}$ (it is
$\Lambda _L^{6-2(N_f-1)}=m^2\Lambda ^{6-2N_f}$ for $N_c=3$).
Similarly, the matching relation associated with breaking $SO(N_c)$
with $N_f$ vectors to $SO(N_c-1)$ with $N_f-1$ vectors by an
expectation value $\ev{Q^{N_f}}$ is $\Lambda
_L^{3(N_c-2)-N_f-2}=\Lambda ^{3(N_c-2)-N_f}(M^{N_fN_f})^{-1}$ (for
breaking $SO(4)\cong SU(2)_1\times SU(2)_2$ to $SO(3)$, it is $\Lambda
_L^{6-2(N_f-1)}= 4\Lambda _1^{6-N_f}\Lambda _2^{6-N_f}
(M^{N_fN_f})^{-2}$).}.

The electric theory has an anomaly free global $SU(N_f)\times U(1)_R$
symmetry with the fields $Q$ transforming as
$(N_f)_{(N_f-N_c+2)/N_f}$.  The dual theory has the same global
symmetry with $M$ transforming as $Q\cdot Q$ and $q$ transforming as
$(\overline N_f)_{(N_c-2)/N_f}$, which is anomaly free and respected
by the superpotential \typsup.  Furthermore, the 'tHooft anomalies of
this magnetic spectrum match those of the electric spectrum.

For $N_c-2< N_f \le {3 \over 2} (N_c-2)$, the magnetic degrees of
freedom are free in the infra-red while, for ${3
\over 2} (N_c-2)< N_f < 3 (N_c-2)$, the electric and the magnetic
theories flow to the same non-trivial fixed point of the
renormalization group. Although the two theories are different away
{}from the extreme infra-red, they are completely equivalent at long
distance.  This means that the two (super) conformal field theories at
long distance are identical, having the same correlation functions of
all of the operators, including high dimension (irrelevant) operators.

The gauge invariant (primary) chiral operators of the electric theory
are
\eqn\sonop{\eqalign{
M^{ ij }&= Q^iQ^j \cr
B^{[i_1,...,i_{N_c}]}&=Q^{i_1}...Q^{i_{N_c}} \cr
b^{[i_1,...,i_{N_c-4}]}&=W_\alpha ^2 Q^{i_1}...Q^{i_{N_c-4}} \cr
\CW_\alpha^{[i_1,...,i_{N_c-2}]}&=W_\alpha Q^{i_1}...Q^{i_{N_c-2}} \cr
}}
with the gauge indices implicit and contracted.  These operators get
mapped to gauge invariant operators of the magnetic theory as
\eqn\emopmap{\eqalign{M^{ij}&\rightarrow M^{ij}\cr
B^{[i_1\dots i_{N_c}]}&\rightarrow \epsilon ^{i_{1}\dots i_{N_f}}
\widetilde b _{i_{N_c+1}\dots i_{N_f}}\cr b^{[i_1\dots
i_{N_c-4}]}&\rightarrow
\epsilon ^{i_1\dots i_{N_f}}\widetilde B_{[i_{N_c-3}\dots i_{N_f}]}\cr
\CW_\alpha^{[i_1,...,i_{N_c-2}]}&\rightarrow \epsilon ^{i_1\dots
i_{N_f}}(\widetilde{\CW}_\alpha)_{[i_{N_c-1},...,i_{N_f}]},\cr}} where
$\widetilde B$, $\widetilde b$, and $\widetilde\CW_\alpha$ are the
magnetic analogs of the operators in \sonop.  These maps are
compatible with the global symmetries discussed above.

Deforming the electric theory along a flat direction of non-zero
$\ev{Q^{N_f}}$ gives $SO(N_c-1)$ with $N_f-1$ massless flavors
and the new fixed point is at weaker coupling.  In the dual
description this deformation gives a mass $\mu ^{-1}\ev{M^{N_fN_f}}$ to
$q_{N_f}$ and the low energy dual theory is $SO(N_f-N_c+4)$ with
$N_f-1$ massless quark flavors, which is the dual of the low energy
electric theory.  In the magnetic description the new fixed point is
at stronger coupling and the relation \mgscrgenn\ is preserved in the
low energy theories.

Some classical identities of the electric theory arise from quantum
effects in the dual.  For example, it is seen classically in the
electric theory that rank$(M)\leq N_c$.  In the magnetic description,
this condition arises because, for larger rank, a superpotential is
dynamically generated and the magnetic theory has no vacuum.
Similarly, the electric theory has a classical relation that, when
rank$(M)=N_c$, the baryon $B^{[i_1, \dots i_{N_c}]}$ has the non-zero
eigenvalue $B=\pm
\sqrt{\det 'M}$.  This is mapped under duality to the relation that,
for $N_f=N_c-4$, the operator $b$ defined above is given by
$b=\pm\sqrt{\Lambda ^{2(N_c-1)}}$, which is related to gaugino
condensation.

Deforming the electric theory by giving a mass to $Q^{N_f}$ gives
$SO(N_c)$ with $N_f-1$ massless flavors and the new fixed point is at
stronger coupling.  In the dual description, adding $W_{\rm tree}
=\half mM^{N_fN_f}$ to \typsup\ and integrating out the massive fields
gives $\ev{q_{N_f}\cdot q_{N_f}}=-\mu m$ which, along with the
$D$-terms, implies that the dual gauge group is broken by the Higgs
mechanism to $SO(N_f-N_c+3)$ with $N_f-1$ massless flavors, which is
the dual of the low energy electric theory.  In the magnetic
description the new fixed point is at weaker coupling and the relation
\mgscrgenn\ is preserved in the low energy theories.

For $N_f=N_c-1$ the dual gauge group is $SO(3)$ and the
superpotential \typsup\ is modified to
\eqn\wmncmi{W={1\over 2\mu}M^{ij}q_i\cdot q_i-{1\over 64
\Lambda _{N_c, N_c-1}^{2N_c-5}}\det M.}
Also, the relation \mgscrgenn\ between the scales is modified in this
case to
\eqn\mgiiisc{2^{14}(\Lambda _{N_c, N_c-1}^{2N_c-5})^2\widetilde
\Lambda _{3, N_c-1}^{6-2(N_c-1)}=\mu ^{2(N_c-1)}.}
These modifications arise upon
going from $N_f=N_c$ with a mass term added for $Q^{N_c}$
to the low energy theory with $N_f=N_c-1$ because of
peculiarities associated
with the breaking of the magnetic $SO(4)\cong SU(2)\times SU(2)$ to
the diagonally embedded magnetic $SO(3)$ \refs{\sem , \isson}.

\newsec{Abelian Coulomb phase}

\subsec{General features}
Consider a general theory with a low energy $N=1$ Abelian Coulomb
phase.  For simplicity we consider only the situation with a single
photon.  As discussed in sect.\ 2, the effective gauge coupling $\tau
_{\rm eff}(X_r,g_I)$ depends holomorphically on the light fields $X_r$
and the coupling constants, including the scale $\Lambda$ of the
underlying non-Abelian $G$ theory in which the low energy, Abelian
theory is embedded.  However, $\tau _{\rm eff}$ is not a single valued
function of $X_r$ and the couplings.  This is possible because $\tau$
gives a redundant description of the physics: $\tau$ is physically
identified under $SL(2,Z)$ transformations, generated by $S:\tau
\rightarrow -1/\tau$, which is associated with the possibility of
exchanging electric with magnetic in ordinary Maxwell theory, and
$T:\tau \rightarrow \tau +1$, which is a unit shift of the theta
angle.  In order for physics to be single valued, $\tau$ need only be
a section of an $SL(2,Z)$ bundle \swi.

For simplicity, we will consider the case where $\tau$ only depends on
a single light field (or a single function of the light fields)
$U$ whose expectation value serves as an order
parameter for breaking the underlying non-Abelian $G$ gauge theory
to a $U(1)$ subgroup.  For large $U$, the $G$ gauge
theory is weakly coupled and the one loop beta function in the
microscopic theory leads to
\eqn\asymptau{\tau \approx {i b \over 2\pi} \log {U\over \Lambda^p,}}
for some integers $b$ and $p$, where $\Lambda$ is the scale of $G$.
As we circle around infinity, $U \rightarrow e^{2\pi i} U$, $\tau
\rightarrow \tau - b$; i.e.\ $\tau$ is transformed by
$\CM_\infty=T^{-b}$.  So even at weak coupling $\tau$ is not
single valued.  The low energy effective gauge coupling ${1 \over
g_{\rm eff}^2} \sim \Im \tau$ is invariant under $\CM _\infty$.
However, if $\Im \tau$ were single valued everywhere in the interior
of the moduli space, because it is a harmonic function, it couldn't be
everywhere positive definite \swi.  There would then be regions in the
moduli space where $g_{\rm eff}$ is imaginary.  This unphysical
conclusion can be avoided if the topology of the moduli space is
complicated in the interior or, as found in
\swi, there are several (at least two) singular values $U_i$ of $U$
with monodromies $\CM_i$ around them which do not commute with
$\CM_\infty= T^{-b}$.

The monodromies $\CM_i$ around the $U_i$ must have a physical
interpretation.  The simplest one is that they are associated with
$k_i$ massless particles at the singularity.  The low energy
superpotential near $U_i$ then has the form
\eqn\lowengw{W_L{}^{(i)} =  (U-U_i) \sum_{l=1}^{k_i} c_l^{(i)}
\widetilde
E_l^{(i)} E_l^{(i)} + \CO ((U-U_i)^2)} where
$\widetilde E_l^{(i)}$ and
$E_l^{(i)}$ are the new massless states.  If the constants $c_l^{(i)}$
are nonzero, these states acquire a mass of order $ \CO ((U-U_i)) $
away from the singularity.  Therefore, the one loop beta function in
the low energy theory leads to
\eqn\lowbet{\tau_i \approx -{ i k_i \over 2 \pi} \log (U-U_i) }
(we assume for simplicity that, as in \refs{\swi, \swii},
all the $E_l^{(i)}$ have
charge one; the generalization to other cases is straightforward)
where $\tau_i$ is the coupling to the low energy photon.  $\tau_i$ is
related to $\tau$ in the asymptotic region by a duality transformation
$N_i$.  It is clear from \lowbet\ that the monodromy in $\tau_i$ is
$T^{k_i}$.  Therefore, the monodromy in $\tau$ is
\eqn\monti{\CM_i = N_i^{-1} T^{k_i} N_i .}
For $\CM_i$ to not commute with $\CM_\infty= T^{-b}$, the
transformation $N_i$ must include $S$.  This means that the massless
particles $E_l^{(i)}$ at $U_i$ are magnetically charged.

As discussed in \swi, because $\tau$ is a section of an
$SL(2,\bZ)$ bundle it is naturally described as the modular parameter
$\tau$ of a torus.
A torus is conveniently described by the one complex
dimensional curve in $C^2$:
\eqn\torusc{y^2=x^3+Ax^2+Bx+C,}
where $(x,y)\in C^2$ and $A$, $B$ and $C$ are parameters.  The modular
parameter of the torus \torusc\ is given by
\eqn\taucrv{\tau (A,B,C)={\int _b{dx\over y}\over \int _a{dx\over
y}},}
where $a$ and $b$ refer to a basis of cycles around the branch cuts of
\torusc\ in
the $x$ plane.  The problem of finding the section $\tau$ is thus
reduced to the simpler problem of finding $A$, $B$, and $C$ as
functions (rather than sections) of $U$ and the various coupling
constants and scales.

The
$\tau$ obtained from \torusc\
is singular when the torus is singular, which is when
\eqn\toruss{x^3+Ax^2+Bx+C=0\qquad\hbox{and}\qquad 3x^2+2Ax+B=0.}
Eliminating $x$, this is when the discriminant of the cubic equation in
\toruss\ vanishes: $\Delta (A,B,C)=0$ where
\eqn\singabc{\Delta=4A^3C-B^2A^2-18ABC+4B^3+27 C^2.}
As discussed in \refs{\swi, \swii} , the order of the zero can be used
to determine the monodromy \monti\ around the singularity and, thus,
the charge of the associated massless fields.

Important constraints \swii\ on the dependence of the coefficients in
\torusc\ on $U$ and the coupling constants are the following:
\lfm{1.} In the weak coupling limit $\Lambda =0$ the curve should be
singular for every $U$.  Without loss of generality we can then take
$y_0^2=x^2(x-U)$.
\lfm{2.} The parameters $A,\ B,\ C$ in \torusc\ are holomorphic in $U$
and the various coupling constants.  This guarantees that $\tau$ is
holomorphic in them.
\lfm{3.} The curve \torusc\ must be compatible with all the global
symmetries of the theory including those which are explicitly broken by
the coupling constants or the anomaly.
\lfm{4.} In various limits
(e.g.\ as some mass goes to zero or infinity)
we should recover the curves of other models.
\lfm{5.} The curve should have physical monodromies around the
singular points.

\subsec{$SO(N_c)$ with $N_f=N_c-2$}

For $N_f=N_c-2$ the gauge group is broken by $\ev{Q}$ to $SO(2)
\cong U(1)$ and the theory has an Abelian Coulomb phase.
Applying the considerations of the previous section, it is found that
the Abelian Coulomb phase has an effective gauge coupling $\tau _{\rm
eff}(\det M, \Lambda)$ which is exactly given by the curve \isson\
\eqn\curve{y^2=x^3+x^2(-\det M+8\Lambda ^{2N_c-4})+
16\Lambda ^{4N_c-8}x.}
For example,
at weak coupling (large $\det\ev{M}$), \curve\ properly reproduces the
one loop beta function of $SO(N_c)$ with $N_f$ fields $Q$.
The curve \curve\ has singularities at the solutions of
\singabc, which are $\det M=0$ and $\det M=16\Lambda ^{2N_c-4}$.

Classically, the submanifold $\det M=0$ has a singularity associated
with a non-Abelian Coulomb phase with some of the $SO(N_c)$ gluons
becoming massless.  In the quantum theory, the monodromy of $\tau$
implied by \curve\ around $\det M=0$ reveals that this submanifold of
the moduli space of vacua is actually in an Abelian free magnetic
phase associated with massless monopoles.  At the origin there are
$N_f$ massless monopoles $q_i^\pm$.  Away from the origin, they obtain
a mass matrix proportional to $\ev{M}$ via
\eqn\tausaysi{W\sim M^{ij}q_i^+q_j^-.}

The monodromy obtained from \curve\ around $\det M=16\Lambda
^{2N_c-4}$ reveals that this submanifold is in a free dyonic phase
associated with a single dyonically charged field $E$ which is
massless at $\det M=16\Lambda ^{2N_c-4}$; near the singularity the
dyon gets a mass via
\eqn\tausaysii{W\sim (\det M-16\Lambda ^{2N_c-4})E^+E^-.}
The quantum numbers of the dyon
$E$ are such that $E\sim q_iQ^i$.  For example, if we take the $q_i$ to
be monopoles with zero electric charge, $E$ is a dyon with
electric charge one.  Thus, as a matter of convention, we will refer to
the $q_i$ as monopoles and to $E$ as a dyon.

To summarize, we should expect to find two dual descriptions of the
original electric theory, one in which the
monopoles $q_i$ are taken as
fundamental fields and one in which the dyon $E$ is
a taken as a fundamental field.

Consider the duality discussed in the previous section.  For
$N_f=N_c-2$ the dual magnetic theory is $SO(2)\cong U(1)$ with $N_f$
charged fields $q^\pm _i$ and neutral fields $M^{ij}$ with a
superpotential
\eqn\wmoniim{W\sim M^{ij}q_i^+q_j^-.}
To see the relation of this Abelian dual to the non-Abelian duals
considered in the previous section, consider flowing from $N_f=N_c-1$
to $N_f=N_c-2$ by giving an electric quark a mass.  In the magnetic
theory this generically Higgses the magnetic $SO(3)$ to a magnetic
$SO(2)$ and the low energy superpotential is \wmoniim.  (There are
additional contributions to \wmoniim\ from instantons in the broken
magnetic $SO(3)$ \isson.)  The components $q_i^\pm$ of the magnetic
quarks become magnetic monopoles in the low energy Abelian Coulomb
phase.  The $N_f$ monopoles $q_i^\pm$ with superpotential \wmoniim\ is
precisely the situation \tausaysi, determined above from  the curve
\curve.

The dyon \tausaysii\ is seen by a strong coupling analysis in either
the electric or the magnetic theories \refs{\sem , \isson}.  There is
also a dyonic dual description in which the dyons $E$ and $\widetilde
E$ are taken as fundamental fields.  Just as the $N_f=N_c-2$, Abelian
magnetic dual can be obtained by flowing down from the non-Abelian,
$N_f=N_c-1$, $SU(2)$ magnetic dual, the Abelian dyonic dual can be
obtained by flowing down from a dyonic dual, $N_f=N_c-1$, $SO(N_c)$
gauge theory.  This dyonic dual theory
\isson\ has composite $SO(N_c)$ gauge
fields interacting with a theta angle which differs from that of the
original electric theory by $\pi$ and it has composite quarks $d^i$,
satisfying $d^i\cdot d^j=M^{ij}$, with superpotential
\eqn\wddual{W=-{1\over 32\Lambda ^{2N_c-5}}\det d\cdot d.}
Adding $W_{tree}=\half mQ^{N_c-1}\cdot Q^{N_c-1}
=\half md^{N_c-1}\cdot d^{N_c-1}$ the low energy electric theory has
$N_f=N_c-2$ and the low energy dyonic dual has an unbroken
$SO(2)$ with a charged pair which becomes massless at
$\det M=16\Lambda _L^{2N_c-4}$, coming
{}from $d^{N_c-1}$, which is seen at weak coupling.

Consider giving a mass to the $N_c-2$-th flavor by adding $W_{\rm
tree}=\half mM^{N_c-2,N_c-2}$, giving $SO(N_c)$ with $N_f=N_c-3$ light
flavors at low energy.  Adding $W_{\rm tree}$ eliminates the Coulomb
phase and the low energy theory has two distinct branches.  One branch
is associated with adding $W_{\rm tree}$ to \wmoniim\ which, upon
integrating out the massive fields, gives the monopole condensate
$\ev{q_{N_f}q_{N_f}}\sim m$, yielding a confining phase by the dual
Meissner effect.  The low energy theory has a moduli space of vacua
labeled by $\ev{\widehat M}$ with $N_c-3$ massless fields at the
origin, with a superpotential given by
\eqn\exoticw{W\sim M^{ij}q_iq_j,}
where we dropped the hats and $i,j=1\dots N_c-3$.  In the magnetic
description, the fields $q_i$ are left-over components of the magnetic
quarks.  By the map \emopmap, the quark component $q_i$ in this case
is mapped to the exotic $q_i\sim (Q)^{N_c-4}W_\alpha W^\alpha$.
Intuitively, one thinks of such exotics as being large and heavy bound
states.  Here we see that they become massless at $\ev{M}=0$.  This
phenomenon is similar to the massless composite mesons and baryons
found in $SU(N_c)$ with $N_f=N_c+1$ \nati.  As was the case there, we
have confinement without chiral symmetry breaking.

The other branch is associated with adding $W_{\rm tree}$ to
\tausaysii.  Integrating out the massive fields gives the dyon
condensate $\ev{E^+E^-}\sim m/\det
\widehat M$, yielding an oblique
confining phase.  The low energy theory has a
superpotential
\eqn\wobcon{W_{\rm oblique}={8\Lambda _L^{2N_c-3}
\over \det \widehat M}.}
In the electric description, this superpotential is associated with
gaugino condensation in the unbroken electric $SO(3)$ along with
contributions from instantons in the broken $SO(N_c)/SO(3)$ \isson.
In the magnetic description
\wobcon\ arises from dyon condensation.  In the dyonic description
\wddual\ it arises at tree level.

\newsec{$SO(3)\cong SU(2)$ examples}

The discussion of the previous sections is modified slightly
for $SO(3)\cong SU(2)$, where the $3$ dimensional representation
is the adjoint representation.

\subsec{One adjoint, $Q$; an Abelian Coulomb phase}

This is the $N=2$ theory discussed in \swi.  The theory has a quantum
moduli space of vacua labeled by the expectation value of the massless
meson field $M= Q^2$.  The $SU(2)$ gauge symmetry is broken to $U(1)$
on this moduli space, so the theory has a Coulomb phase with a
massless photon.

As discussed in \refs{\swi,\swii}, the effective
gauge coupling in the Coulomb phase
is given by the $\tau _{\rm eff}(M)$ obtained from the curve
\eqn\niicrv{y^2=x^3-Mx^2+4\Lambda ^4x.}
(This curve is expressed using the convention for the normalization of
$\tau$ discussed in \swii, $\tau={\theta\over \pi}+8\pi ig^{-2}$.)
This gives a $\tau _{\rm eff}(M)$ which
has singularities associated with a
massless magnetic monopole field
$q_{(+)}$, at $M=4\Lambda ^2$ and a massless dyon $q_{(-)}$ at\foot{We
use the conventions of
\refs{\ils, \finnpou} where the normalization of $\Lambda^2$ (in the
$\overline{DR}$ scheme) differs by a factor of 2 from that of \swi; our
order parameter $M$ is related to $u$ of \swi\ as $u=\half M$.}
$M=-4\Lambda ^2$.  Therefore, these two points are in a free magnetic
and a free dyonic phase, respectively.  Here $q_{(+)}$ is a doublet
charged under the magnetic $U(1)_M$, which is related to the electric
$U(1)_E$ by the electric-magnetic transformation $S$: $F\rightarrow
\widetilde F$ (modulo $\Gamma(4) \subset SL(2,Z)$).
Similarly, $q_{(-)}$ is
a doublet charged under a dyonic $U(1)_D$, related to $U(1)_E$ by the
$SL(2,Z)$ transformation $ST^2$ (again, modulo $\Gamma(4) \subset
SL(2,Z)$), where $T$ is a rotation of the theta angle by $\pi$.  Near
where these fields are massless, they couple through the effective
superpotentials
\eqn\mdwii{W_\pm \sim  \left(M  \mp 4\Lambda ^2\right)
q_{(\pm )}\cdot q_{(\pm )}.}

Referring to the underlying $SU(2)$ theory as ``electric,'' we can say
that it has two dual theories.  One of them, which we can refer to as
the ``magnetic dual,'' describes the physics around $M=4\Lambda^2$ with
the superpotential $W_+$.  The other dual, which can be called the
``dyonic dual,'' is valid around $M=-4\Lambda^2$ and is described by
$W_-$.

Consider giving $Q$ a mass by adding a term $W_{\rm tree}= \half mM$
in the electric theory.  Adding $W_{\rm tree}$ to \mdwii, the
equations of motion give $\ev{q_{(\pm)}\cdot q_{(\pm)}}\sim m$ and
lock $\ev{M}=\pm 4\Lambda ^2$.  The condensate of monopoles/dyons
Higgses the dual theory and thus gives confinement/oblique confinement
of the electric theory by the dual Meissner effect \swi.

Consider analyzing, as in sect.\ 2.3, the 1PI effective action for
this theory with sources.  Starting from the analog of
\ymw\ for this theory,
$W_L(m)=\pm 2(\Lambda ^4m^2)^{1/2}$, equations
\corr\ and \invlt\ give $W=0$ with
the constraint $\ev{M}=\pm 4\Lambda ^2$.  Indeed, adding the source
$m$ for $M$ drives the theory to the confining or oblique confining
phase with $\ev{M}=\pm 4\Lambda ^2$. The Coulomb phase cannot be
explored in the theory with a mass term for $Q$.  As discussed in
sect.\ 2.3, this method of analyzing the theory must fail to capture
some of the physics because the theory without sources has massless
fields (the monopole or dyon) which cannot be
represented by the gauge invariant observables.

Because this theory actually has $N=2$ supersymmetry,
it can be further analyzed using
the additional techniques applicable for $N=2$ theories,
yielding the Kahler potential for $Q$ and a BPS mass bound \swi.

\subsec{$SU(2)$ with two adjoints; A non-Abelian Coulomb phase}

This theory has $N=1$ (not $N=2$) supersymmetry.  Writing the matter
fields as $Q^i$ with $i=1,2$ a flavor index, there is a 3 complex
dimensional moduli space of classical vacua parameterized by the
expectation values of the gauge singlet fields $M^{ij}=Q^i\cdot Q^i$.
In the generic vacuum $\ev{Q^1}$ breaks $SU(2)$ to a $U(1)$ which is
then broken by $\ev{Q^2}$.  For $\det \ev{M^{ij}}\neq 0$, the gauge
group is completely broken and the theory is in the Higgs phase.  On
the non-compact two complex dimensional subspace of vacua with $\det
M=0$, there is an unbroken $U(1)$ gauge symmetry and thus a light
photon along with a pair of massless electrically charged fields. At
the point $\ev{M}=0$ the $SU(2)$ gauge group is unbroken.

We now turn to the quantum theory.  The theory has the global symmetry
group $SU(2)\times U(1)_R$, with $Q$ transforming as ${\bf 2}_{\half}$,
which determines that any dynamically generated superpotential must be
of the form
\eqn\wnfiiwsym{W={c\over \Lambda}\det M,}
with $c$ a dimensionless constant.  Its behavior at $M\rightarrow
\infty$ is incompatible with asymptotic freedom, as signaled by the
presence of the scale $\Lambda$ in the denominator.  Therefore, no
superpotential can be generated and the classical vacuum degeneracy
outlined above is not lifted quantum mechanically.

The generic ground state with generic $M$ is in the Higgs phase.
Consider now the subspace of the moduli space with $\det M=0$.  The
low energy degrees of freedom there are a single photon, a pair of
massless electrically charged fields and some neutral fields.  This
theory cannot become strong in the infrared.  In fact, the loops of
the massless charged fields renormalize the electric charge to zero.
Therefore, this subspace of the moduli space is in a free electric
phase.

Now consider adding a tree level superpotential $W_{\rm tree}=\half
\Tr mM$.  Taking $m=\pmatrix{0&0\cr 0&m_{2}}$, $Q^2$ gets a mass and
can be integrated out.  The low energy theory is $SU(2)$ with a single
massless adjoint matter field, which is the example of the previous
subsection.  This low energy theory has a scale $\Lambda _L$, which is
related to that of the original theory by
$\Lambda_L^4 = m_2^2 \Lambda^2$, and
a massless monopole or dyon at
$\ev{M^{11}}=\pm 4\Lambda _L^2=\pm 4m_2\Lambda$.
Note that as
$m_2\rightarrow 0$ the point $\ev{M}=0$ has both massless monopoles
and dyons.  These are mutually non-local\foot{ A similar situation was
found in $N=2$ $SU(3)$ Yang Mills theory
\ref\AD{P. Argyres and M. Douglas, hep-th/9505062, IASSNS-HEP-95-31.}.
} and signal another phase at this point in the theory with $m_2=0$.
We interpret this as a non-Abelian Coulomb phase \intse.

Starting from the theory with $m_2\neq 0$, turning on $m_1\neq 0$
drives the monopole or dyon to condense and the vacuum is locked at
$\ev{M^{11}}=\pm 4m_2\Lambda$.  The $+$ sign is a vacuum with monopole
condensation and thus confinement.  The $-$ sign is a vacuum with dyon
condensation and thus oblique confinement.  More generally, these
vacua are at $\ev{M^{ij}}=\pm 4\Lambda \det m(m^{-1})^{ij}$.  These
expectation values can be obtained from
\eqn\we{W_e={e\over 8\Lambda}\det M+\half \Tr mM,}
with $e=\mp 1$ for confinement and oblique confinement, respectively.

The theory has various phase branches.  For mass $m=0$ there is a
Higgs phase which, in terms of $W_e$, corresponds to $e=0$.  There is
a subspace $\det M=0$ in the free electric phase and the point $M=0$
in a non-Abelian Coulomb phase.  For $m\neq 0$ but with $\det m=0$ the
theory is in the Coulomb phase with a free magnetic point and a free
dyonic point.  For $\det m\neq 0$ the theory is either confining and
described by the superpotential \we\ with $e=-1$ or it is oblique
confining and described by the superpotential \we\ with $e=1$.

If we consider the 1PI
effective action, the analog of \ymw\ is $W_L(m)=\pm 2(\Lambda ^2\det
m)^{1/2}$.  Integrating in gives the confining or oblique confining
phase branches of the superpotential \we,
with $e=\pm 1$, missing the $e=0$ Higgs phase branch.
Again, as discussed in sect.\ 2.3, the 1PI superpotential
necessarily fails to capture
some of the physics because the theory without the sources has
massless particles, the quarks and the gluons, which cannot be
represented by the gauge invariant observables.

The analysis of \isson\ reveals that this electric theory has two dual
descriptions, similar to the magnetic and dyonic duals discussed in
the previous section, labeled by $\epsilon =\pm 1$.  The two dual
theories are based on an $SU(2)$ gauge group with two fields $q_i$ in
its adjoint representation and three gauge singlet fields $M^{ij}$.
The difference between the two theories is in the superpotential
\eqn\wepe{W_{\epsilon} ={1\over 12\sqrt{\Lambda \widetilde \Lambda}}
M^{ij}q_i\cdot q_j+\epsilon \left({1\over 24\Lambda}\det M+{1\over
24\widetilde
\Lambda}\det q_i\cdot q_j\right),}
where $\widetilde \Lambda$ is the scale of the dual $SU(2)$ (we
expressed $\mu$ in terms of $\Lambda$ and $\widetilde \Lambda$).  The
theory with $\epsilon =1$ is a ``magnetic'' dual and that with
$\epsilon =-1$ a ``dyonic'' dual.

We now analyze the dynamics of these dual theories.  Since they are
similar to the original electric theory, we proceed
as we did there. These theories have three phases: Higgs, confining
and oblique confining.  We study them using the gauge invariant order
parameters $N_{ij}\equiv q_i\cdot q_j$.  Its effective superpotential
is obtained by writing the tree level superpotential \wepe\ in terms
of $N$ and adding to it ${\tilde e \over 8 \widetilde \Lambda} \det N$
where, in the Higgs, confining and oblique confinement branches,
$\tilde e= 0, -1, 1$, respectively
\eqn\wepen{W_{\epsilon, \tilde e} ={1\over 12\sqrt{\Lambda \widetilde
\Lambda}}\Tr MN+\epsilon \left({1\over 24\Lambda}
\det M+{1\over 24\widetilde
\Lambda}\det N\right)+{\tilde e\over 8\widetilde \Lambda}\det N.}
Now we can integrate out the massive field $N$ to find
\eqn\wepenon{W_{\rm eff} ={1\over 8\Lambda}{\tilde e-\epsilon\over
1+3\tilde e\epsilon}\det M.}
This is the same as the effective superpotential \we\ of the electric
theory with
\eqn\etran{e={\tilde e-\epsilon\over 1+3\tilde e\epsilon}.}
We see that the various phases are permuted in the different
descriptions as:
\smallskip
$$\vbox{\rm \settabs 4 \columns
\+ {\bf Theory}\qquad\qquad\qquad &\ &{\bf Phases}&\ \cr
\+ &\ &\ &\ \cr
\+ electric & Higgs ($e=0$) & conf. ($e=-1$) & obl. conf. ($e=1$) \cr
\+ magnetic ($\epsilon=1$) & obl.
conf. ($\tilde e=1$) &  Higgs ($\tilde
e=0$ ) & conf. ($\tilde e=-1$) \cr
\+ dyonic ($\epsilon =-1$)
& conf. ($\tilde e=-1$) & obl. conf. ($\tilde
e=1$) & Higgs ($\tilde e=0$) \cr} $$

\smallskip

It is a simple exercise to check that by dualizing the magnetic and
dyonic theories as we above dualized the electric theory (two duals of
each), we find permutations of the same three theories. The $S_3$
triality permuting the phases and branches is associated with a
quotient of the $SL(2,Z)$ electric-magnetic duality symmetry group:
the theories are preserved under $\Gamma (2)\subset SL(2,Z)$, leaving
the quotient $S_3=SL(2,Z)/\Gamma (2)$ with a non-trivial action.

This discussion leads to a new interpretation of the first term in \we
.  In the electric theory this term appears as a consequence of
complicated strong coupling dynamics in the confining and the oblique
confinement branches of the theory.  In the dual descriptions it is
already present at tree level.

Consider the theory with a mass $m_2$ for $Q^2$. As discussed above,
the low energy electric theory has a Coulomb phase with massless
monopoles or dyons at the strong coupling singularities
$\ev{M^{11}}=\pm 4m_2\Lambda$.  We now derive this result in the dual
theories.  Adding $W_{\rm tree}=\half m_2M^{22}$ to the superpotential
\wepe\ of the dual theory, the equations of motion give
\eqn\dceom{\eqalign{{1 \over 12\sqrt{\Lambda \widetilde \Lambda}
} q_2 \cdot q_2 +{8\epsilon \over 24\Lambda} M^{11}+{1\over 2}m_2&=0
\cr q_1 \cdot q_2&=0}\qquad
\eqalign{M^{22}&=-\half \epsilon\sqrt{\Lambda \over \widetilde \Lambda}
q_1 \cdot q_1 \cr M^{12}&=0.}}  For $q_2^2\neq 0$, $\ev{q_2}$ breaks
the gauge group to $U(1)$ and the remaining charged fields $q_1^{\pm}$
couple through the low energy superpotential
\eqn\dcloww{{1\over 16\sqrt{\Lambda \widetilde \Lambda}}
(M^{11}-4\epsilon m_2 \Lambda)q_1^+q_1^-.}  (This superpotential is
corrected by contributions from instantons in the broken magnetic
$SU(2)$ theory.  However, these are negligible near $M^{11}=4\epsilon
m_2 \Lambda$.)  We see that the theory has a charged doublet of
massless fields $q_1^\pm$ at $M^{11}=4\epsilon m_2\Lambda$, exactly as
expected from the analysis of the electric theory.  There these states
appeared as a result of strong coupling effects.  Here we see them as
weakly coupled states in the dual theories.  This is in accord with
the interpretation of the $\epsilon=1$ ($\epsilon=-1$) theory as
magnetic (dyonic).

The other monopole point on the moduli space of the theory with
$m_1=0$ but $m_2\not=0$ is at $M^{11}=-4\epsilon m_2\Lambda$.  It
arises from strong coupling dynamics in the dual theories.  To see
that, note that the above analysis is not valid when the expectation
value of $q_2$ is on the order of or smaller than the mass of $q_1$.
In that case, $q_1$ should be integrated out first.  The equations of
motion in the low energy theory yield a single massless monopole point
at $M^{11}=-4\epsilon m_2\Lambda $ \isson.

An analysis similar to the one above leads to a strongly coupled
state in the dual theories along the flat directions with $\det M=0$
in the $m=0$ case.  This state can be interpreted as the massless
quark of the electric theory in that free electric phase.

To conclude, this theory has three branches which are in three
different phases: Higgs, confining and oblique confinement (various
submanifolds of these branches are in Coulomb, free electric, free
magnetic and free dyonic phases).  They touch each other at a point in
a non-Abelian Coulomb phase.  Corresponding to the three branches
there are three different Lagrangian descriptions of the theory:
electric, magnetic and dyonic.  Each of them describes the physics of
one of the branches, where it is Higgsed, in weak coupling and the
other two in strong coupling.

In both examples of this section, the theory has a discrete symmetry
which relates the confining and the oblique confinement
phases\foot{This symmetry is manifest only in the electric
description.  In the dual descriptions it is realized as a quantum
symmetry \isson.}.  Therefore, in these cases the effects of
confinement are indistinguishable from the effects of oblique
confinement.  Correspondingly, the magnetic and the dyonic
descriptions are similar -- they differ only in the sign of
$\epsilon$.  In the other $SO(N_c)$ examples discussed in
the previous section, these two phases are not related by a symmetry
and the two dual descriptions look totally different.

\newsec{Conclusions}

To conclude, supersymmetric field theories are tractable and many of
their observables can be computed exactly\foot{Although we did not
discuss them here, we would like to point out that many other
examples were studied
\nref\mos{A.Yu. Morozov, M.A. Olshansetsky and M.A. Shifman,
\np{304}{1988}{291}.}%
\nref\iss{K. Intriligator, N. Seiberg and S. Shenker, hep-ph/9410203,
\pl {342}{1995}{152}.}%
\nref\sunnt{A. Klemm, W. Lerche, S. Theisen and S. Yankielowicz,
hep-th/9411048, \pl{344}{1995}{169}; hep-th/9412158.}%
\nref\arfa{P. Argyres and A. Faraggi, hep-th/9411057, \prl{74}{1995}
{3931}.}%
\nref\kutasov{D. Kutasov, hep-th/9503086, \pl{351}{1995}{230}.}%
\nref\rlms{R. Leigh and M. Strassler, hep-th/9503121,
\np{447}{95}{1995}}%
\nref\dansun{U. Danielsson and B. Sundborg, USITP-95-06, UUITP-4/95,
hep-th/9504102.}%
\nref\doush{M.R. Douglas and S.H. Shenker, RU-95-12, RU-95-12,
hep-th/9503163.}%
\nref\jerus{S. Elitzur, A Forge, A. Giveon and E. Rabinovici, RI-4-95
hep-th/9504080.}%
\nref\asy{O. Aharony, J. Sonnenschein and S. Yankielowicz,
TAUP--2246--95, CERN-TH/95--91, hep-th/9504113.}%
\nref\schkut{D. Kutasov and A. Schwimmer, EFI--95--20, WIS/4/95,
hep-th/9505004.}%
\nref\intpou{K. Intriligator and P. Pouliot, hep-th/9505006,
\pl{353}{1995}{471}.}%
\nref\intdual{K. Intriligator,
RU--95--27, hep-th/9505051, Nucl. Phys. B
to appear.}%
\nref\ad{P.C. Argyres and M.R. Douglas, RU-95-31, hep-th/9505062.}%
\nref\berkooz{M. Berkooz, RU-95-29, hep-th/9505067.}%
\nref\rlmsspso{R. Leigh and M. Strassler, hep-th/9505088, RU-95-30.}%
\nref\ilst{K. Intriligator, R. Leigh and M. Strassler, hep-th/9506148,
RU-95-38.}%
\nref\poul{P. Pouliot, hep-th/9507018, RU-95-46}%
\nref\pesando{I. Pesando, hep-th/9506139, NORDITA-95/42 P}%
\nref\giddpie{S.B. Giddings and J. M Pierre,
hep-th/9506196, UCSBTH-95-14}%
\nref\poptriv{E. Poppitz and S. P. Trivedi, hep-th/9507169, EFI-95-44,
Fermilab-Pub-95/258-T}
\refs{\ads, \cern, \mos, \ils, \swii, \intse, \sem, \iss-\poptriv}
exhibiting many new interesting phenomena.}.  Our analysis led us to
find new phases of non-Abelian gauge theories, like the non-Abelian
Coulomb phase with its quantum equivalence and the free magnetic phase
with its massless composite gauge fields.
The main dynamical lesson is the role of electric-magnetic
duality in non-Abelian gauge theories in four dimensions.  To be clear
we should distinguish several different notions of duality:

\item{1.} The exact
$\tau \rightarrow -1/\tau$ duality of Maxwell theory and its
generalization to the Montonen-Olive \mo\ duality of finite,
interacting, non-Abelian theories.

\item{2.} The $\tau _{eff}\rightarrow -1/\tau _{eff}$ duality of low
energy theories with an Abelian Coulomb phase.  This is not a
symmetry but, rather, an ambiguity in the description of the low
energy physics.

\item{3.} The
duality of two asymptotically free theories which flow to the same
non-Abelian Coulomb phase fixed point in the infrared.

\item{4.} The duality of the free magnetic phase, which provides a
relation between the UV and
the IR behavior of a theory which is free in the IR.

\noindent
There are relations between many of the phenomena discussed above
\refs{\isson, \rlms} but, at a deeper level,
they remain to be really understood.

\bigskip
\centerline{{\bf Acknowledgments}}

We would like to thank T. Banks, D. Kutasov, R. Leigh, M.R.  Plesser,
P.  Pouliot, S.  Shenker, M. Strassler and especially E. Witten for
many helpful discussions.  This work was supported in part by DOE
grant \#DE-FG05-90ER40559.

\listrefs
\end